\documentclass[twocolumn,openany,openright,amsmath,amssymb,superscriptaddress]{revtex4-2}

\usepackage[english]{babel}
\usepackage{float}
\usepackage{latexsym} 
\usepackage{amsmath,amsthm}
\usepackage{ifpdf}
\usepackage{epstopdf}
\usepackage{subfig}
\usepackage{soul}
\usepackage{dcolumn}% Align table columns on decimal point
\usepackage{bm}% bold math
\usepackage{braket}
\usepackage{wrapfig}
\usepackage[dvipsnames]{xcolor}
\usepackage{color}
\usepackage{hyperref}

\usepackage{graphicx}

\newcommand{\beq}{\begin{equation}}
\newcommand{\eeq}{\end{equation}}
\newcommand{\barr}{\begin{eqnarray}}
\newcommand{\earr}{\end{eqnarray}}
\newcommand{\bseq}{\begin{subequations}}
\newcommand{\eseq}{\end{subequations}}
\newcommand{\vett}[1]{\textbf{#1}}
\newcommand{\uvett}[1]{\hat{\textbf{#1}}}
\usepackage[normalem]{ulem}

\begin{document}

%Single-photon nonlinearity of Epsilon-Near-Zero media in optical nanocavities
\title{Nonlinear Quantum Electrodynamics of Epsilon-Near-Zero Nanocavities}
\author{Luca Dal Negro}\email{Corresponding author: dalnegro@bu.edu}%\thanks{Both authors contributed equally to this work}
\affiliation{Department of Electrical \& Computer Engineering, Boston University, 8 Saint Mary's Street, Boston, 02215, MA, USA}
\affiliation{Department of Physics, Boston University, 
590 Commonwealth Avenue, Boston,02215, MA, USA}
\affiliation{Division of Materials Science \&  Engineering, Boston University, 15 St. Mary’s street, Brookline, 02446, MA, USA}
\author{Riccardo Franchi}
\affiliation{Department of Electrical \& Computer Engineering, Boston University, 8 Saint Mary's Street, Boston, 02215, MA, USA}
\author{Marco Ornigotti}
\affiliation{Faculty of Engineering and Natural Sciences, Tampere University, Tampere, Finland}

\begin{abstract}
	We investigate single-photon nonlinear refractive index change and frequency shift of Epsilon-Near-Zero (ENZ) sub-wavelength nanocavities.
    We apply the rigorous quantum Langevin-noise approach in the framework of Green's tensor quantization method to realistic ENZ materials with causal dispersion and derive closed-form analytical solutions for cavities with spherical geometry.
    This is achieved by employing a fully nonperturbative methodology for the analysis of open quantum systems with single-photon Kerr-type nonlinearity. The analytical results are validated numerically using the established quasi-normal mode expansion method and extended to nonspherical nanocavity geometries that can be experimentally fabricated using state-of-the-art electron lithography.  
    Our findings establish a rigorous benchmark for understanding single-photon nonlinear optical effects in Kerr-type ENZ nanostructures with losses and are of importance to emerging quantum technology applications, including on-chip single-photon nondemolition detection, quantum sensing, and controlled quantum gates driven by enhanced photon blockade effects at the nanoscale.
\end{abstract}

\maketitle

\section{Introduction}
In recent years, the development of optics-based quantum technologies has witnessed an impressive growth driven by the demonstration of single-photon sources and entangled photon pairs that leverage nonlinear optical phenomena such as spontaneous parametric down-conversion (SPDC) or four-wave mixing (SFWM) \cite{o2009photonic,dousse2010ultrabright,tonndorf2017chip,eisaman2011invited,boyer2008entangled,zhang2022spatially}.
Light states with non-classical correlations are important for many applications ranging from quantum communication and cryptography, computing and information processing, to fundamental tests of quantum mechanics. 
In particular, single-photon nonlinear optics on solid-state nanostructures bear the promise to revolutionize quantum information technologies by providing scalable and energy efficient solutions for the engineering of controlled quantum gates, entangled photon sources, novel quantum sensors and quantum nondemolition detection (QND) devices with room-temperature (RT) operation  \cite{eisaman2011invited,munro2005high,flayac2015all,ferretti2012single,PhysRevA.66.063814,imoto1985quantum,xiao2008quantum,munro2005high,braginsky1996quantum}.
However, in most studies of nonlinear sources and detectors, light-matter coupling is considered either semi-classically or within the quantum electrodynamic theory (QED) of linear dissipative systems, such as in the Huttner-Barnett model \cite{huttner1992quantization}.
Specifically, this approach builds upon the Hopfield model \cite{hopfield1958theory} of homogeneous and isotropic bulk dielectrics by introducing a field polarization vector linearly coupled to a continuum distribution of harmonic oscillators representing a reservoir.
Irreversible absorption processes are decribed by an essentially unidirectional energy flow that proceeds from the medium polarization to the reservoir.
In this approach, the polarization field and the harmonic-oscillator ``heat bath'' give rise to dressed-matter operators that combine with the free electromagnetic field to give rise to the quantum mechanical polariton operators of the absorbing medium.
However, while several methods are available to address the QED of linear and dispersive dielectrics  \cite{knoll2000qed,bechler2006path,vogel2006quantum,dung2006,gruner1996green,difallah2019path}, it remains challenging to adequately describe the regime of resonant nonlinear optical interactions in dispersive Epsilon-Near-Zero (ENZ) materials and nanostructures with refractive index dispersion and absorption losses that satisfy the Kramers-Kronig relation.
In particular, in strongly absorbing media the familiar field expansion obtained in terms of propagating modes fails completely and simple alternatives based on non-orthogonal evanescent waves are incomplete. 
Recently, significant advances have been achieved towards the QED of nonlinear absorbing media using the Langevin-noise approach in combination with the Green's dyadic framework of macroscopic electrodynamics, motivating our present contribution  \cite{crosse2010effective,scheel2006causal,scheel2006quantum, krstic2023nonperturbative}.

In this paper, we use the quantum Langevin-noise
approach in the framework of Green's tensor quantization method to analytically compute the effective Kerr-type nonlinear phase shift and refractive index change of dispersive ENZ materials in spherical nanocavities.  Moroever, we validate our predictions using numerical analysis of quasi-normal modes \cite{ren2021quasinormal,lalanne2019quasinormal,kristensen2020modeling} and extend the study to more general structures with nonspherical shapes.
In particular, we focus on dispersive ENZ photonic structures with the goal of quantifying single-photon nonlinear optical effects at the nanoscale in materials that offer large and nonperturbative Kerr-type effective nonlinearity \cite{alam2016large,capretti2015enhanced,reshef2017beyond,reshef2019nonlinear,shubitidze2024enhanced,tamashevich2024field,shubitidze2025enhancement,capretti2015comparative}.
Our findings provide an analytic framework to rigorously assess whether a single infrared photon can nonlinearly perturb the refractive index of the ENZ polariton medium at the nanoscale, potentially controlling the propagation of a probe photon.
Currently, single-photon nonlinear optical effects have been demonstrated only using Rydberg atoms in ultra-high Q cavities, which suffers from limited bandwidth, cryogenic operation temperature, and difficult integration with photodetectors \cite{chang2014quantum}.
Therefore, there is a need to investigate alternative approaches that leverage the large Kerr-type effective nonlinearity of nanoscale optical cavities in order to achieve QND detection in a solid-state platform.

Our paper is organized as follows: in section \ref{estimates} we begin by motivating our analysis based on a preliminary estimate of the nonlinear phase change in a non-dispersive Kerr medium embedded in a photonic cavity at the single photon level.
Section \ref{linear} and \ref{nonlinear} briefly review the quantization approach for linear and Kerr-type nonlinear dispersive materials, respectively, which are described by an effective $\chi^{(3)}$ susceptibility, on which our subsequent analysis will be based.
In sections \ref{nanocavity} and \ref{sec:Kerr_freq_shift} we obtain closed-form expressions for the nonlinear refractive index and frequency shift of general ENZ resonators.
In section \ref{sec:Kerr_ENZ_sphere}, we apply our general results to the case of spherical ENZ nonocavities with different radii and derive analytical expressions for their nonlinear Kerr-type responses.
In section \ref{sec:QNMs}, using numerical analysis based on quasinormal modes we provide the path to apply our theoretical results to nanocavities with arbitrary geometry.
Finally, in section \ref{conclusions}, we draw our conclusions.

\section{Preliminary considerations}\label{estimates}
We provide here a preliminary discussion at a qualitative level of the single-photon nonlinearity expected in optical cavities and  establish a simple figure of merit for the development of solid-state cavity quantum electrodynamics (SSQED) in nonlinear dielectric materials. We start by addressing the condition that enables a nonlinear phase shift $\Delta\phi_{\rm NL}$$\sim\pi$ in Kerr-type dielectric media at the single photon level.
Our preliminary estimates neglect optical losses and dispersion effects, which will be treated rigorously in the subsequent sections. 
The feasibility of compact devices for SSQED applications is qualitatively assessed by considering the phase shift induced by light in a Kerr-type medium \cite{PhysRevA.66.063814}:
\begin{equation}\label{eq0}
	\Delta\phi_{\rm NL}=\frac{1}{2}k_{0}L\chi^{(3)}E^{2},
\end{equation}
where $k_{0}$ is the free-space wavenumber, $\chi^{(3)}$ the third-order optical susceptibility of the medium, $L$ is a characteristic interaction length in the material, and $E$ is the local electric field amplitude. 
For a single photon, $E=\sqrt{\hbar\omega/(2\epsilon_{0}V)}$ and $V$ is the volume of the medium that induces the phase shift. Therefore, we can write the nonlinear phase shift due to a single photon field as follows:
\begin{equation}\label{eq2}
	\Delta\phi_{\rm NL}=\frac{\hbar\omega^{2}L\chi^{(3)}}{4c\epsilon_{0}V}.
\end{equation}
The last expression, when evaluated with parameters that are typical of nonlinear dielectric structures (i.e., $\chi^{(3)}$$\sim$$10^{-22} m^{2}/V^{2}$) with macroscopic size (i.e., order of a millimeter in length) yields $\Delta\phi_{\rm NL}$$\sim$$10^{-18}$, which is a negligibly small value for all practical purposes.
This may lead one to conclude that single-photon nonlinear optical effects cannot be achieved in Kerr-type devices.
However, recent developments in highly nonlinear ENZ materials \cite{alam2016large,capretti2015enhanced,reshef2017beyond,reshef2019nonlinear,shubitidze2024enhanced,tamashevich2024field,shubitidze2025enhancement,capretti2015comparative}
and inverse designed high-Q dielectric nanostructures \cite{jensen2011topology,choi2017self,albrechtsen2022nanometer,christiansen2021inverse,mignuzzi2019nanoscale,molesky2018inverse} provide novel opportunities in photonic resonant cavities with sub-wavelength mode confinement and large quality factor $Q=2\pi\nu\tau_{ph}=2\pi\nu/(c\alpha)$, where $\nu$ is the frequency of the cavity, $\tau_{ph}$ is the cavity photon lifetime, and $\alpha$ the overall cavity loss coefficient.
For cavity structures, we can estimate the characteristic photon interaction length $L=1/\alpha=Q/k_{0}$, resulting in:
\begin{equation}\label{eq3}
	\Delta\phi_{\rm NL}=\frac{\hbar\omega Q\chi^{(3)}}{4\epsilon_{0} V}.
\end{equation}

\noindent However, the simple formula in Eq. (\ref{eq3}) is only valid for non-dispersive materials and electromagnetic fields confined in cavity volumes larger than the wavelength. The main objective of this work is to generalize Eq. (\ref{eq3}) to the case of sub-wavelength nanocavities with dispersive nonlinear media, which is accomplished based on Langevin-noise field quantization and resonant states \cite{resStates2} or quasi-normal modes theory \cite{wu2021nanoscale}.
In this context, the large effective nonlinear coefficient $\chi^{(3)}$$\sim$$10^{-16} m^{2}/V^{2}$ of Indium Tin Oxide (ITO) nanolayers  \cite{reshef2017beyond,alam2016large,shubitidze2024enhanced,britton2022structure,tamashevich2024field} and the demonstration of high-$Q$ cavities with nanoscale mode volumes
$V$=$10^{-3}$$\mu m^{3}$ by inverse design and topological optimization methods \cite{albrechtsen2022nanometer,choi2017self,albrechtsen2022nanometer,christiansen2021inverse,mignuzzi2019nanoscale}, makes it important to revisit nonlinear effects at the sub-wavelength scale in order to achieve $\Delta\phi_{\rm NL}$$\approx\pi$ in the single-photon regime on a solid-state platform, requiring a full QED approach.

\section{Field quantization in dispersive and lossy linear media}\label{linear}
In this section, we outline the Langevin-noise approach used for the quantization of the electromagnetic field in linear and locally responding causal media, previosuly developed in Refs. \cite{knoll2000qed,dung1998three,gruner1996green}. 
This method is formulated in terms of the Green's dyadic response of macroscopic electrodynamics and can be naturally applied to the quantization of electromagnetic fields in cavities with arbitrary shapes and dimensions, including sub-wavelength nanocavities, provided the corresponding resonant states can be computed.

We begin by considering the classical Maxwell's equations in the frequency space and in the absence of free charges and currents:
\bseq\label{eq1}
\begin{align}
\nabla\cdot\vett{D}(\vett{r},\omega)&=0,\\
\nabla\cdot\vett{B}(\vett{r},\omega)&=0,\\	\nabla\times\vett{E}(\vett{r},\omega)&=i\omega\vett{B}(\vett{r},\omega),\\
\nabla\times\vett{B}(\vett{r},\omega)&=-i\omega\mu_{0}\vett{D}(\vett{r},\omega),
\end{align}
\eseq
where
\begin{equation}
	\vett{D}(\vett{r},\omega)=\epsilon_{0}\vett{E}(\vett{r},\omega)+\vett{P}(\vett{r},\omega),
\end{equation}
and the polarization of an isotropic medium is defined as
\begin{equation}
	\vett{P}(\vett{r},\omega)=\epsilon_{0}\chi^{(1)}(\vett{r},\omega)\vett{E}(\vett{r},\omega)+\vett{P}_{N}(\vett{r},\omega),
\end{equation}
with the susceptibility $\chi^{(1)}(\vett{r},\omega)$ describing the linear response of the dispersive medium and $\vett{P}_{N}(\vett{r},\omega)$ is the linear noise-polarization field corresponding to the Langevin noise. This term, associated to the presence of absorption losses, must be included in the linear polarization for the theory to be consistent with the fluctuation-dissipation theorem \cite{knoll2000qed,scheel2006quantum,vogel2006quantum,drezet2017equivalence,drezet2017quantizing,krstic2023nonperturbative}.
Due to this noise-polarization field, the spectral Fourier components of the electric field obey the noise-driven inhomogeneous Helmholtz equation:
\begin{equation}
	\nabla\times\nabla\times\vett{E}(\vett{r},\omega)-\frac{\omega^{2}}{c^{2}}\varepsilon(\vett{r},\omega)\vett{E}(\vett{r},\omega)=\omega^{2}\mu_{0}\vett{P}_{N}(\vett{r},\omega),
\end{equation}
where we introduced the complex (relative) permittivity of the medium $\varepsilon(\vett{r},\omega)=\varepsilon_1(\vett{r},\omega)+i\varepsilon_2(\vett{r},\omega)=1+\chi^{(1)}(\vett{r},\omega)$.
The equation above can be readily solved using the dyadic Green's function $\overleftrightarrow{\vett{G}}(\vett{r},\vett{r}^{\prime},\omega)$ for the vector wave equation, yielding: 
\begin{equation}
	\vett{E}(\vett{r},\omega)=\omega^{2}\mu_{0}\int_{V}d^{3}\vett{r}^{\prime}
	\overleftrightarrow{\vett{G}}(\vett{r},\vett{r}^{\prime},\omega)\cdot
	\vett{P}_{N}(\vett{r}^{\prime},\omega),
\end{equation}
where the dyadic Green's function satisfies
\beq\label{eqGreen1}
\nabla\times\nabla\times\overleftrightarrow{\vett{G}}(\vett{r},\vett{r}^{\prime},\omega)-\frac{\omega^2}{c^2}\varepsilon(\vett{r},\omega)\overleftrightarrow{\vett{G}}(\vett{r},\vett{r}^{\prime},\omega)={\delta}(\vett{r}-\vett{y}^{\prime})\overleftrightarrow{\mathbb{I}},
\eeq
and $\overleftrightarrow{\mathbb{I}}$ denotes the unit dyadic.

The electromagnetic field quantization in the dispersive material is achieved by introducing the local bosonic field operators $\hat{\vett{f}}\,(\vett{r},\omega)$ and $\hat{\vett{f}}^{\dag}(\vett{y},\omega)$ that describe collective microscopic excitations of the electromagnetic field and the linear absorbing material.
The bosonic operators generalize the free-space photonic mode operators to the case of absorptive dielectric media and satisfy the equal-time canonical commutation relations $[\hat{f}_{\mu}(\vett{r},\omega),\hat{f}^{\dag}_{\nu}(\vett{r}^{\prime},\omega)]=\delta_{\mu\nu}\delta(\omega-\omega^{\prime})\delta(\vett{r}-\vett{r}^{\prime})$.
Moreover, the noise-polarization field is related to the bosonic operators by the relation
\cite{knoll2000qed,scheel2006quantum,vogel2006quantum,drezet2017equivalence,drezet2017quantizing}
\begin{equation}
	\vett{P}_{N}(\vett{r},\omega)=i\sqrt{\frac{\hbar\epsilon_{0}}{\pi}\epsilon_2(\vett{r},\omega)}\,\hat{\vett{f}}(\vett{r},\omega).
\end{equation}
This allows us to obtain, in the Heisenberg picture, the frequency components of the quantized electric field in the medium in terms of the canonically conjugate polariton-like dynamical variables \cite{gruner1996green,dung1998three,knoll2000qed,scheel2006causal,scheel2006quantum,vogel2006quantum}:
\beq\label{eq11}
\hat{\vett{E}}(\vett{r},\omega)=i\sqrt{\frac{\hbar}{\pi\varepsilon_0}}\frac{\omega^2}{c^2}\int\,d^3\vett{r}^{\prime}\,\sqrt{\varepsilon_2(\vett{r}^{\prime},\omega)}\,\overleftrightarrow{\vett{G}}(\vett{r},\vett{r}^{\prime},\omega)\hat{\vett{f}}(\vett{r}^{\prime},\omega).
\eeq
The expression above establishes the linear response between the noise operators and the electric field and shows that the classical dyadic propagator $\overleftrightarrow{\vett{G}}(\vett{r},\vett{r}^{\prime},\omega)$ is the response function of the problem.
Consequently, a direct application of the fluctuation-dissipation theorem establishes the fundamental link between the zero-temperature vacuum fluctuations of the electric field and the imaginary part of the Green's function \cite{knoll2000qed,scheel2006causal, krstic2023nonperturbative}:
\begin{equation}\label{eq12a}
	\langle{0}|\hat{\vett{E}}(\vett{r},\omega)
	\hat{\vett{E}}^{\dag}(\vett{r}^{\prime},\omega^{\prime})|0\rangle=\frac{\hbar\omega^{2}}{\pi\epsilon_{0}c^{2}}\text{Im}[\vett{G}(\vett{r},\vett{r}^{\prime},\omega)]\delta(\omega-\omega^{\prime}).
\end{equation}

The total electric field operator is then defined as the analytic signal
\beq
\hat{\vett{E}}(\vett{r})=\int_0^{\infty}\,d\omega\,\hat{\vett{E}}(\vett{r},\omega)+\text{h.c.},
\eeq
Finally, the dynamic operator-valued Maxwell's equations are generated from the bilinear Hamiltonian
\beq\label{eq15a}
\hat{H}_0=\int\,d^3r\,\int_0^{\infty}\,d\omega\,\hbar\omega\,\hat{\vett{f}}^{\dagger}(\vett{r},\omega)\hat{\vett{f}}(\vett{r},\omega).
\eeq
The Hamiltonian above is also the generator of the time evolution for the noise operators $\hat{\vett{f}}(\vett{x},\omega)$, which is governed by the Heisenberg equation of motion
\beq\label{eq16}
\frac{\partial}{\partial t}\hat{\vett{f}}(\vett{r},\omega; t)=\frac{1}{i\hbar}\left[\hat{\vett{f}}(\vett{r},\omega; t),\hat{H}_0\right],
\eeq
Since the electric field operator is linearly defined in terms of noise operators through Eq. \eqref{eq11}, the same equation of motion also describes the quantum dynamics of the field operator.

\section{Nonlinear Quantum Electrodynamics of dispersive media}\label{nonlinear}

To investigate the quantum dynamics of the electromagnetic field in a dispersive, absorbing, and Kerr-type nonlinear medium within the Green's dyadic quantisation approach introduced above, one can define the following interaction Hamiltonian 
\barr\label{eq18}
\hat{H}_{Kerr}&=&\int\,[d^3r\,d\omega]_4\alpha_{\mu\nu\sigma\tau}([\vett{r},\omega]_4)\hat{f}_{\mu}(\vett{r}_1,\omega_1)\nonumber\\
&\times&\hat{f}_{\nu}(\vett{r}_2,\omega_2)\hat{f}^{\dagger}_{\sigma}(\vett{r}_3,\omega_3)\hat{f}^{\dagger}_{\tau}(\vett{r}_4,\omega_4)+\text{h.c.},
\earr
where $[\vett{r},\omega]_N=(\vett{r}_1,\cdots,\vett{r}_N,\omega_1,\cdots,\tilde\omega_n)$, $[d^3r\,d\omega]_N=\prod_{i=1}^Nd^3r_id\omega_i$, and $\alpha_{\mu\nu\sigma\tau}([\vett{r},\omega]_4)$ is the polaritonic nonlinear (Kerr) coupling coefficient describing the Kerr interaction at the microscopic level in terms of the nonlinear electromagnetic interaction mediated by the matter. Notice, that the quantity $\alpha_{\mu\nu\sigma\tau}$ is not the usual third-order susceptibility $\chi^{(3)}_{\mu\nu\sigma\tau}$, typical of macroscopic nonlinear optics, but is its microscopic counterpart defined for the dressed light-matter field. As discussed in detail in Refs. \cite{gruner1996green,dung1998three,knoll2000qed,scheel2006causal,scheel2006quantum,vogel2006quantum}, this quantity is not directly accessible experimentally, since the noise operators $\hat{f}_{\mu}(\vett{r},\omega)$ do not correspond to physical observables. Therefore, in order to achieve a complete description of nonlinear interactions at the microscopic level, one should connect  $\overleftrightarrow{\alpha}([\vett{x},\omega]_4)$ with the usual third-order susceptibility tensor used in the macroscopic description $\overleftrightarrow{\chi}^{(3)}(\vett{x},[\omega]_4)$ \cite{boyd_nonlinear_2020}. In general, however, there is no simple relation between these two quantities.
The only regime in which a connection has been established is within the slowly varying amplitude approximation (SVAA), where the electromagnetic field envelope is assumed to vary on a much slower timescale compared to its carrier frequency \cite{crosse2010effective,scheel2006causal}. In this approximation, an explicit analytic expression relating $\overleftrightarrow{\alpha}([\vett{x},\omega]_4)$ and $\overleftrightarrow{\chi}^{(3)}(\vett{x},[\omega]_4)$ can be established \cite{scheel2006quantum,scheel2006causal}, and the quantisation scheme defined above can be used in combination with the standard expressions for the nonlinear macroscopic polarisation in dispersive and absorbing media.

Specifically, for a Kerr-type nonlinear interaction, the SVAA fields can be considered to oscillate at four different frequencies $\{\Omega,\Omega_1,\Omega_2,\Omega_3\}$, with the constraint that $\Omega=\Omega_1+\Omega_2-\Omega_3$ holds, which encapsulates energy conservation between the interacting frequency modes \cite{boyd_nonlinear_2020}.
The SVAA electric field operator can then be written as
\beq\label{eq24}
\hat{E}_{\mu}(\vett{r},t)=\hat{\mathcal{E}}_{\mu}(\vett{r},t;\Omega)e^{-i\Omega t}+\sum_{k=1}^3\hat{\mathcal{E}}_{\mu}(\vett{r},t;\Omega_k)e^{-i\Omega_kt},
\eeq
where $\hat{\mathcal{E}}_{\mu}(\vett{r},t;\Omega)$ has the form of Eq. \eqref{eq11}, with the bosonic operators $\hat{\vett{f}}(\vett{r},t,\omega)$ replaced by the SVAA bosonoic operators 
\beq\label{eqHoperator}
\hat{\vett{h}}(\vett{r},t;\Omega_k)=\frac{1}{\sqrt{\Delta\Omega_k}}\int_{\Delta\Omega_k}d\omega\,\hat{\vett{f}}(\vett{r},t,\omega),
\eeq
where $\Delta\Omega_k$ is the spectral width of the SVAA signal oscillating at carrier frequency $\Omega_k$, and $\Delta\Omega_k\ll\Omega_k$ holds. 

The nonlinear polarisation operator can be then constructed starting from the standard definition of the nonlinear third-order polarization \cite{wilhelmi}
\barr
\hat{P}^{(NL)} _{\mu}(\vett{r},t)&=&3\varepsilon_0\int_{-\infty}^t\,[d\tau]_3\,\chi^{(3)}_{\mu\beta\sigma\tau}(\vett{r},t-\tau_1,t-\tau_2,t-\tau_3)\nonumber\\
&\times&\hat{E}_{\beta}(\vett{r},\tau_1)\hat{E}_{\sigma}(\vett{r},\tau_2)\hat{E}^{\dagger}_{\tau}(\vett{r},\tau_3)+\text{h.c.},
\earr
and using the SVAA approximation, i.e., Eq. \eqref{eq24} for the electric field operator to take out the electric field amplitudes from the integrals above, and Eqs. \eqref{eq11} and \eqref{eqHoperator} to construct the electric field operator. This allows us to write the slowly-varying, macroscopic, quantum nonlinear polarisation operator as

\barr\label{eq29}
\hat{P}^{(NL)}_{\mu}(\vett{r};\Omega)&=&3\varepsilon_0\chi^{(3)}_{\mu\beta\sigma\tau}(\vett{r},\Omega_1,\Omega_2,-\Omega_3)\nonumber\\
&\times&\hat{\mathcal{E}}_{\beta}(\vett{r},\Omega_1)\hat{\mathcal{E}}_{\sigma}(\vett{r},\Omega_2)\hat{\mathcal{E}}^{\dagger}_{\tau}(\vett{r},\Omega_3)+\text{h.c.},
\earr
where the time dependence of the operators above has been suppressed for convenience.

This result allows one to construct an effective Kerr Hamiltonian from the microscopic one defined in Eq. \eqref{eq18}, which reads
\barr\label{eq30}
\hat{H}_{Kerr}^{(eff)}&=&\frac{3\varepsilon_0}{2}\int\,d^3r\,\chi^{(3)}_{\mu\nu\sigma\tau}(\vett{r},\Omega_1,\Omega_2,\Omega_3)\Bigg[\nonumber\\
&\times&\hat{\mathcal{E}}_{\mu}(\vett{r},\Omega_1)\hat{\mathcal{E}}_{\nu}(\vett{r},\Omega_2)\hat{\mathcal{E}}_{\sigma}^{\dagger}(\vett{r},\Omega_3)\hat{\mathcal{E}}_{\tau}(\vett{r},\Omega_4)^{\dagger}\nonumber\\
&+&\text{h.c.}\Bigg],
\earr
together with the third-order energy conservation constraint $\Omega_1+\Omega_2=\Omega_3+\Omega_4$. Unlike the Hamiltonian \eqref{eq18}, this Hamiltonian contains experimentally measurable quantities, like susceptibilities and electric fields, rather than inaccessible microscopic quantities, like polaritonic nonlinear coupling constants and polaritonic operators. The effective Hamiltonian above, therefore, can be used to extract information about the quantum nonlinear dynamics of a dispersive and absorbing medium, as, for example, the Kerr-induced refractive index change in a nanocavity, as we will show in the next Section.

\section{Kerr-type nonlinear index change of ENZ nanocavities}\label{nanocavity}
We now consider a geneal ENZ nanocavity, characterized by a volume $V$ and with an arbitrary shape.
Recalling that for the Kerr effect $\Omega_1=\Omega_2=\Omega_3=\Omega$ holds \cite{boyd_nonlinear_2020}, the displacement operator, constructed using $\vett{D}=\vett{D}^{(L)}+\vett{P}^{(NL)}+\vett{P}^{(N)}$, %and \eqref{eq29}, 
becomes
\barr
\hat{D}_{\mu}(\vett{r},\Omega)&=&\varepsilon_0\varepsilon(\vett{r},\Omega)\hat{\mathcal{E}}_{\mu}(\vett{r},\Omega)\nonumber\\
&+&3\varepsilon_0\chi^{(3)}_{\mu\beta\sigma\tau}(\vett{r},\Omega)\hat{\mathcal{E}}_{\beta}(\vett{r},\Omega)\hat{\mathcal{E}}_{\sigma}(\vett{r},\Omega)\hat{\mathcal{E}}^{\dagger}_{\tau}(\vett{r},\Omega).
\earr
where $\varepsilon(\vett{r},\Omega)$ is the relative dielectric permittivity of the medium. Notice, that all the fields appearing above are slowly varying fields. From the expression above, one can extract the field-corrected permittivity $\varepsilon^{(fc)}_{\mu\beta}$ by writing $\hat{D}_{\mu}=\varepsilon_{\mu\beta}^{(fc)}(\vett{r},\omega)\hat{E}_{\beta}$ and taking the expectation value of the displacement operator over a suitable quantum state $\ket{\psi}$, representing the state of the electromagnetic field interacting with the nonlinear medium. By applying Wick's theorem \cite{srednicki} to transform the trilinear term $\langle\hat{E}\hat{E}\hat{E}^{\dagger}\rangle$ into $\langle\hat{E}\hat{E}^{\dagger}\rangle\langle\hat{E}\rangle$, one obtains
\barr\label{eq32}
\varepsilon^{(fc)}_{\mu\beta}(\vett{r},\Omega)&=&\varepsilon(\vett{r},\Omega)\delta_{\mu\beta}\nonumber\\
&+&3\chi^{(3)}_{\mu\beta\sigma\tau}(\vett{r},\Omega)\langle\hat{E}_{\sigma}(\vett{r},\Omega)\hat{E}^{\dagger}_{\tau}(\vett{r},\Omega)\rangle,
\earr
which is in agreement with the standard results from nonlinear optics. For a single mode field, one can remove the indices from the expression above and obtain the usual expression of the Kerr permittivity, which is proportional to the field intensity \cite{boyd_nonlinear_2020}. 

The Kerr-induced nonlinear refractive index change per unit volume inside the ENZ resonator can be calculated as follows
\beq\label{eq34}
\Delta n_{\mu\nu}=\frac{1}{V}\int_V\,d^3r\,\left[\sqrt{\varepsilon^{(fc)}_{\mu\nu}(\vett{r},\Omega)}-\sqrt{\varepsilon(\vett{r},\Omega)\delta_{\mu\nu}}\right],
\eeq
This definition is consistent with the nonperturbative approach of ENZ nonlinearities introduced by Reshef and Boyd \cite{reshef2017beyond}, where $\sqrt{\varepsilon_{\mu\nu}^{(fc)}(\vett{r},\Omega)}$ is the nonperturbative refractive index. Equations \eqref{eq32} and \eqref{eq34} are generally valid and constitute an important result of our work, as they yield the fully quantum nonlinear refractive index change due to the Kerr effect.

A related quantity, which contains the same information as the $\Delta n$ defined above, is the so-called Kerr frequency shift per photon \cite{haroche,kirchmair}, which can be interpreted as the coupling constant of the Kerr Hamiltonian divided by $\hbar$ \cite{drummondKerr}.
In the following section, therefore, starting from the Kerr Hamiltonian in Eq. \eqref{eq30}, we derive the general expression of the Kerr frequency shift and then address the special case of a single photon created in a given mode of the ENZ nanocavity.

\section{Single-photon Kerr Shift }\label{sec:Kerr_freq_shift}
To obtain the expression of the Kerr frequency shift, we need to rewrite the Kerr Hamiltonian in Eq. \eqref{eq30} in the form $\hat{H}_{Kerr}=\hbar\,\Gamma\,\hat{a}\,\hat{a}\,\hat{a}^{\dagger}\,\hat{a}^{\dagger}$, i.e., we need to convert it to Fock space.
Then, we can simply read the Kerr frequency shift as the parameter $\Gamma$.
To represent the Kerr Hamiltonian in terms of Fock space (i.e., photon) operators, we need a suitable relation between the (SVAA) polaritonic operator $\hat{h}_{\mu}(\vett{r})$ and the correspondent Fock operator $\hat{a}$.
To do so, we can use the standard approach of choosing a set of modes to represent the electromagnetic field, and writing the Fock space operators in terms of such modes \cite{birulaPhotons,breuerQS}.
However, due to the presence of losses in the material, we cannot use normal modes anymore, but instead we need the quasi-normal modes (QNMs) \cite{kristensen2020modeling} or resonant states (RSs) \cite{resStates}.
For the spherical geometry, we choose the RSs, as they are easier to handle analytically. The results we show below do not depend, however, on the particular representation of the modes.

Following Ref. \cite{resStates2}, we introduce a complete set of the quasi-normal modes for a lossy resonator, which we call $\vett{E}_n(\vett{r})$, as the eigenstates of the complex eigenvalue problem
\beq
\nabla\times\nabla\times\vett{E}_n(\vett{r})-k^{2}\varepsilon(\tilde{\omega}_n)\vett{E}_n(\vett{r})=0,
\eeq
where $k=\tilde{\omega}/c$ is the wave vector of the electromagnetic field determined by the frequency $\tilde\omega$, which is in general complex. In particular, for a resonant open system with losses we have $\tilde\omega_n=\omega_n-i\gamma_n$ denoting the complex frequency of the resonant states (RSs), where the imaginary part describes the leakage/loss rate \cite{resStates,lobanov2018resonant}.
These are the eigensolutions of the source-free vector Helmholtz equation with outgoing waves boundary conditions at infinity, also known as quasi-normal modes \cite{kristensen2020modeling,sauvan2022normalization,lalanne2018light,lalanne2019quasinormal}, and are understood as generalizations of the familiar concept of the normal modes of Hermitian (closed, non-lossy) systems. These RSs form a complete orthonormal basis for any lossy resonator, with the closure relation \cite{resStates}
\beq\label{closure}
\sum_n\varepsilon(\tilde\omega_n)E_{\mu,n}(\vett{r})E_{\nu,n}(\vett{r}')=\delta_{\mu\nu}\delta(\vett{r}-\vett{r}'),
\eeq
and the following normalisation condition \cite{resStates}
\barr\label{normalisation}
1&=&2\frac{d[\omega^2\varepsilon(\omega)]}{d(\omega^2)}\Bigg|_{\tilde\omega_n}\int_V\,d^3r\,\vett{E}_n(\vett{r})\cdot\vett{E}_n(\vett{r})\nonumber\\
&+&\frac{c^2}{\tilde\omega_n^2}\int_{\partial V}\,d^2r\,\Bigg[\vett{E}_n(\vett{r})\cdot\frac{\partial}{\partial s}\left(\vett{r}\cdot\nabla\right)\vett{E}_n(\vett{r})\nonumber\\
&-&\left(\vett{r}\cdot\nabla\right)\vett{E}_n(\vett{r})\cdot\frac{\partial\vett{E}_n(\vett{r})}{\partial s}\Bigg],
\earr
where $s$ is a coordinate along the direction normal to the surface $\partial V$. Notice, moreover, that contrary to the usual normal modes, where $\vett{E}_n^*(\vett{r})$ enters in the closure and orthogonality relation as the dual of $\vett{E}_n(\vett{r})$, this is not true anymore for RSs, as only $\vett{E}_n(\vett{r})$ appears both in Eq. \eqref{closure} and \eqref{normalisation}.

Moreover, since the RSs form a complete set, they can be used to expand the Green's tensor as follows \cite{resStates,resStates2}
\beq\label{greenRS}
G_{\mu\nu}(\vett{r},\vett{r}',\Omega)=\sum_n\frac{c^2E_{\mu,n}(\vett{r})E_{\nu,n}(\vett{r}')}{\tilde\Omega_n(\Omega-\tilde\Omega_n)}.
\eeq
With these tools, we can then introduce the following relation between the (SVAA) polaritonic operators $\{\hat{h}_{\mu}(\vett{r}),\hat{h}_{\mu}^{\dagger}(\vett{r})\}$ and the Fock space operators $\{\hat{a}_n,\hat{a}^{\dagger}_n\}$, which create or annihilate a photon in a given mode $\vett{E}_n(\vett{r})$, as follows
\bseq\label{hmu}
\begin{align}
\hat{h}_{\mu}^{\dagger}(\vett{r})=\sum_n\,\sqrt{\varepsilon(\tilde\omega_n)}E_{\mu,n}(\vett{r})\hat{a}_n^{\dagger},\\
\hat{h}_{\mu}(\vett{r})=\sum_n\,\sqrt{\varepsilon(\tilde\omega_n)}E_{\mu,n}(\vett{r})\hat{a}_n.
\end{align}
\eseq
Substituting Eqs. \eqref{hmu} and \eqref{greenRS} into Eq. \eqref{eq11} and using the SVAA approximation, we can write the electric field operator in Fock space as
\barr\label{electricRS}
\hat{E}_{\mu}(\vett{r},\Omega)&=&-i\sqrt{\frac{\hbar\Delta\Omega\varepsilon_2(\omega)}{\pi\varepsilon_0}}\nonumber\\
&\times&\sum_n\frac{\Omega^2}{\tilde\omega_n(\Omega-\tilde\omega_n)}E_{\mu,n}(\vett{r})\,\hat{b}_n,
\earr
where $\hat{b}_n\equiv\sqrt{\varepsilon(\tilde\omega_n)}I_n\hat{a}_n$, and
\beq
I_n=\int\,d^3r\,E_{\mu,n}(\vett{r})E_{\mu,n}(\vett{r}).
\eeq
Notice, that the above expression for the electric field operator generalises the usual mode expansion expression for conservative cavities to the case of a nanocavity sustaining RSs, each characterised by its own resonant frequency $\tilde\omega_n$.

Using Eq. \eqref{electricRS} for the field operator, we can then rewrite, after some straightforward algebra, the Kerr Hamiltonian in Eq. \eqref{eq30} as follows
\barr\label{eqKerr2}
\hat{H}_{Kerr}&=&\sum_n\hbar\,\Gamma_n\,\hat{b}_n\,\hat{b}_n\,\hat{b}^{\dagger}_n\,\hat{b}^{\dagger}_n
+\text{h.c.},
\earr
where $\Gamma_n$ is the sought-after Kerr frequency shift per photon, whose general expression is given as follows
\barr\label{eqGamma}
\Gamma_n&=&F_n(\Omega)\int\,d^3r\,\chi_{\mu\nu\sigma\tau}^{(3)}(\vett{r},\Omega)\Bigg[\nonumber\\
&\times&E_{\mu,n}(\vett{r})E_{\nu,n}(\vett{r})E_{\sigma,n}^*(\vett{r})E_{\tau,n}^*(\vett{r})\Bigg],
\earr
where
\barr
F_n(\Omega)&=&\left(\frac{3\hbar\Omega^8\Delta\Omega^2\varepsilon_2^2(\Omega)}{2\pi^2\varepsilon_0}\right)\nonumber\\
&\times&\frac{1}{|\tilde\omega_n(\omega-\tilde\omega_n)|^4}.
\earr

\section{Kerr Shift of ENZ Nanopheres}\label{sec:Kerr_ENZ_sphere}
We now apply the general results obtained in the previous section to a spherical ENZ nanocavity of radius $R$, embedded in a surrounding medium (that we take as air for convenience) and quantify the magnitude of the single photon nonlinear phase shift. The cavity is excited by a slow varying optical pulse with central frequency $\Omega=\omega$ and spectral width $\Delta\Omega=\Delta\omega$. Moreover, we assume that both the nanocavity and the surrounding medium have a homogeneous relative permittivity, so that
\beq\label{eq49}
\varepsilon(\vett{r},\omega)=\begin{cases}
\varepsilon_b-\frac{\omega_p^2}{\omega^2+i\gamma\omega}, &r\leq R,\\
1, & r>R,
\end{cases}
\eeq
where $\varepsilon_b$, $\omega_p$, and $\gamma$ are the residual relative permittivity, plasma frequency, and loss factor, respectively, of the ENZ cavity modeled by the Drude-Sommerfeld model.
As an ENZ material, we consider here ITO and set $\varepsilon_b=3.8$, $\omega_p=3\times 10^{15}$ Hz, and $\gamma=1.91\times 10^{14}$ Hz.
This choice of parameters corresponds to a ENZ wavelength for ITO of $\lambda=1246$ nm.
Moreover, we choose $\chi^{(3)}=5.2 \times 10^{-17}$ $m^2/V^2$ as the value for the effective third-order susceptibility of ITO, which was experimentally measured in Ref. \cite{shubitidze2024enhanced}.
Moreover, we consider $R=10\,{\rm nm}-100\,{\rm nm}$ for the radius of the spherical ENZ nanocavity, and assume a constant SVAA bandwidth of $\Delta\omega=10^{13}\,{\rm Hz}$.

In spherical symmetry, the RSs can be written in terms of the eigenmodes of the spherical cavity as \cite{resStates} $\vett{E}_n(\vett{r})=-\vett{r}\times\nabla\psi_n(\vett{r})$ for TE modes, $\vett{E}_n(\vett{r})=(i/\varepsilon(\omega)k)\nabla\times[-\vett{r}\times\nabla\psi_n(\vett{r})]$ for TM modes, and $\vett{E}_n(\vett{r})=-\nabla\psi_n(\vett{r})$ for the (static) longitudinal electric (LE) modes, where $\psi(\vett{r})$ is a solution of the scalar Helmholtz equation in spherical coordinates \cite{jackson2021classical} and the resonances $\tilde\omega_n$ are found by solving an appropriate secular equation derived from applying the boundary conditions at the cavity surface \cite{resStates}.

Without loss of generality, we can restrict our analysis to TM modes. The same approach can then be used for TE and LE modes as well. Following Ref. \cite{resStates2}, the TM RSs are given by
\barr\label{eq52}
\vett{E}_n(\vett{r})&=&\frac{A_{\ell}^{\rm TM}(\tilde\omega_n)}{k(\tilde\omega_n)r}\Bigg\{\ell(\ell+1)Z_{\ell}(r,\tilde\omega_n)Y_{\ell}^m(\theta,\varphi)\uvett{r}\nonumber\\
&+&\frac{\partial}{\partial r}\left(r Z_{\ell}(r,\tilde\omega_n)\right)\Bigg[\frac{\partial}{\partial\theta}Y_{\ell}^m(\theta,\varphi)\hat{\boldsymbol\theta}\nonumber\\
&+&\frac{1}{\sin\theta}\frac{\partial}{\partial\varphi}Y_{\ell}^m(\theta,\varphi)\hat{\boldsymbol\varphi}\Bigg]\Bigg\},
\earr
where $Y_{\ell}^m(\theta,\varphi)$ are the real-valued spherical harmonics defined as \cite{resStates}
\bseq\label{eq52bis}
\begin{align}
    Y_{\ell}^m(\theta,\varphi)&=\sqrt{\frac{(2\ell+1)(\ell-m)!}{2\pi(\ell+m)!}}P_{\ell}^m(\cos\theta)\chi_m(\varphi),\\
    \chi_m(\varphi)&=\Bigg\{\begin{array}{lr}
    \sin(m\varphi) & m<0,\\
    \frac{1}{\sqrt{2}} & m=0,\\
    \cos(m\varphi) & m>0,
    \end{array}
\end{align}
\eseq
$Z_{\ell}(r,\tilde\omega_n)=j_{\ell}(\sqrt{\varepsilon(\tilde\omega_n)}\tilde\omega_n r/c)/j_{\ell}(\sqrt{\varepsilon(\tilde\omega_n)}\tilde\omega_n R/c)$ inside the nanosphere (with $j_{\ell}(x)$ being the spherical Bessel function of the first kind \cite{nist}) and $Z_{\ell}(r,\omega)=h^{(1)}_{\ell}(\omega r/c)/h^{(1)}_{\ell}(\omega R/c)$ outside the nanopshere (with $h^{(1)}_{\ell}(x)$ being the spherical Hankel function of the first kind \cite{nist}), 
\barr\label{eq53}
A_{\ell}^{\rm TM}(\tilde\omega_n)&=&\sqrt{\frac{1}{\ell(\ell+1)R^3[\varepsilon(\tilde\omega_n)-1]\varepsilon(\tilde\omega_n)D_{\ell}(\tilde\omega_n)}},
\earr
\begin{figure*}[!t]
\begin{center}
   \includegraphics[width=\textwidth]{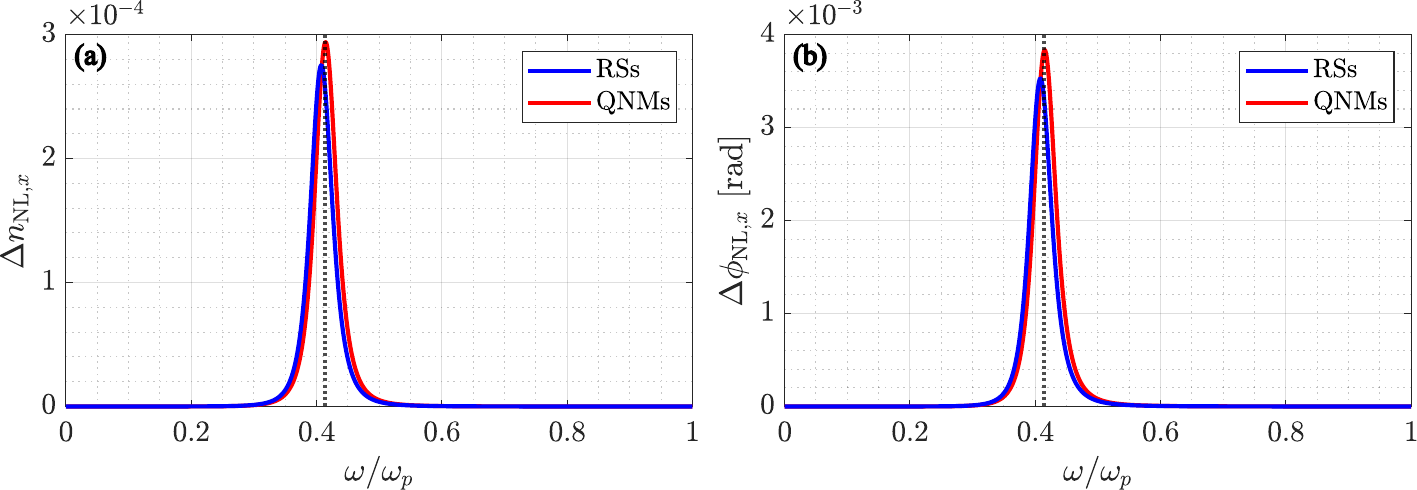}
   \caption{\label{figure1}
   (a) Kerr-induced refractive index change $\Delta n_{\rm NL}$ and (b) Kerr-induced phase shift $\Delta\Phi_{\rm NL}$ as a function of the angular frequency of a single photon within the dipole mode of the sphere ($l=1$ and $m=1$) oriented along $x$.
   The blue curves represent the result evaluated using the RSs, and in red, the ones using the numerically evaluated QNMs.
   The dotted vertical black line identifies the real part of the complex resonant angular frequency calculated through the QNMs ($\omega_1\simeq 0.4138\,\omega_p$).
   In this figure, we consider $R=10\,{\rm nm}$.}
\end{center}
\end{figure*}
where
\bseq
\begin{align}
    D_{\ell}(\omega)&=\frac{1}{\varepsilon(\omega)}\Bigg[\frac{j_{\ell-1}(k(\omega)R)}{j_{\ell}(k(\omega)R)}-\frac{1}{k(\omega)R}\Bigg]^2\nonumber\\
    &+\frac{\ell(\ell+1)}{k^2(\omega)R^2}+\beta_nC_{\ell}(\omega),\\
    C_{\ell}(\omega)&=\frac{1}{\varepsilon(\omega)-1}\Bigg[\frac{j_{\ell-1}^2(k(\omega)R)}{j_{\ell}^2(k(\omega)R)}-\frac{j_{\ell-2}(k(\omega)R)}{j_{\ell}(k(\omega)R)}\nonumber\\
    &-\frac{2l}{k^2(\omega)R^2}\Bigg],\\
    \beta_n&=\frac{\omega}{2\varepsilon(\omega)}\left[\frac{\partial\varepsilon(\omega)}{\partial\omega}\right]_{\omega=\tilde\omega_n},
\end{align}
\eseq
where $R$ is the radius of the cavity, and the resonances $\tilde\omega_n$ for TM polarisation are found from the roots of the following secular equation
\beq
\frac{\sqrt{\varepsilon(\tilde\omega_n)}j_{\ell}'(\sqrt{\varepsilon(\tilde\omega_n)}z)}{j_{\ell}(\sqrt{\varepsilon(\tilde\omega_n)z)}}-\frac{\varepsilon(\tilde\omega_n)h_{\ell}'(z)}{h_{\ell}(z)}-\frac{\varepsilon(\tilde\omega_n)-1}{z}=0,
\eeq
where $z=kR$ and the prime indicates the derivative with respect to the argument of the Bessel and Hankel functions. For small values of $R$, the secular equation above reduces, in the quasi-static limit, to
\beq\label{charEqSmall}
\varepsilon(\omega)=-\frac{\ell+1}{\ell},
\eeq
which is in accordance with the usual results for small spheres \cite{bohren}. Notice, moreover, that the normalisation factor $A^{\rm TM}_{\ell}(\tilde\omega_n)$ defined in Eq. \eqref{eq53} has been obtained by applying the RSs normalisation condition given by Eq. \eqref{normalisation}.

Using the expression for the TM RSs given by Eq. \eqref{eq52} we also calculate the quantity $I_n$, whose explicit expression, for the case $|\tilde\omega_n\sqrt{\varepsilon(\tilde\omega_n)}R/c|\ll 1$, gives
\beq\label{eq57}
I_n=\frac{1}{\tilde\omega_n\varepsilon'(\tilde\omega_n)}.
\eeq
The details of this calculations are reported in Appendix A. We can now apply our general results to the case of a dipole resonance for the ENZ nanosphere.
To do so, we substitute $\ell=1$ and $m=\{0,\pm1\}$ into Eqs. \eqref{eq52} to get, in the quasi-static limit, the following expression for the dipolar RS
\bseq\label{eqDipole}
\begin{align}
    \vett{E}_{1,0}&=\sqrt{\frac{I_1}{V}}\left[\cos\theta\,\uvett{r}-\sin\theta\,\hat{\boldsymbol\theta}\right],\\
    \vett{E}_{1,\pm1}&=\sqrt{\frac{I_1}{V}}\left[-\sin\varphi(\cos\theta\,\uvett{r}+\sin\theta\,\hat{\boldsymbol\theta})\pm\cos\varphi\hat{\boldsymbol\varphi}\right].
\end{align}
\eseq
We then set $\mu=\nu=\sigma=\tau=x$ in Eq. \eqref{eqGamma}, so that $\chi^{(3)}_{\mu\nu\sigma\tau}(\vett{r},\Omega)=\chi^{(3)}_{xxxx}\equiv 5.2\times 10^{-17}$ $m^2/V^2$, as measured experimentally \cite{shubitidze2024enhanced}, and $E_{x,n}(\vett{r})=E_{r,n}(\vett{r})\sin\theta\cos\varphi+E_{\theta,n}(\vett{r})\cos\theta\cos\varphi-E_{\varphi,n}(\vett{r})\sin\theta$. If we then perform the integration over the volume of the nanosphere, we get the following expression for the dipolar Kerr frequency shift
\beq\label{gammaDip}
\Gamma(\omega)=\frac{3\hbar\omega^8\Delta\omega^2\chi^{(3)}(\omega)\varepsilon_2^2(\omega)}{2\pi^2\varepsilon_0 V}\frac{D_{1m}|I_1|^2}{|\tilde\omega_1(\omega-\tilde\omega_1)|^4},
\eeq
where $D_{11}=1$, and $D_{10}=D_{1,-1}=0$. For the simple case of a single mode, uniform, monochromatic field, $\Gamma=\omega\Delta n$ \cite{haroche}, so using this definition we can calculate the Kerr-induced refractive index change as
\beq
\Delta n_{\rm NL}(\omega)\equiv\frac{\Gamma(\omega)}{\omega}=\frac{2\hbar\omega^7\Delta\omega\chi^{(3)}(\omega)\varepsilon_2^2(\omega)}{\pi^2\varepsilon_0 V}\frac{D|I_1|^2}{|\tilde\omega_1(\omega-\tilde\omega_1)|^4}.
\eeq
Analogously, we can calculate the Kerr-induced phase shift by using the relation $\Delta\Phi_{\rm NL}(\omega)=\Gamma(\omega)/(|\gamma_1|)$ (where the modulus is introduced to keep the Kerr phase positive, by convention) \cite{kirchmair} to obtain
\beq\label{DeltaPhiKerr}
\Delta\Phi_{\rm NL}(\omega)\equiv\frac{\Gamma(\omega)}{|\gamma_1|}=\frac{\omega}{|\gamma_1|}\Delta n_{\rm NL}.
\eeq
The Kerr-induced refractive index change and phase shift are plotted in Fig. \ref{figure1}. As it can be seen, both the Kerr-induced refractive index change and phase shift are peaked at the dipolar resonance $\omega=\omega_1$. This is because of the particular choice of the mode (i.e., the dipolar RS) in which the photon is created.
Since the maximum refractive index change is reached at $\omega=\tilde\omega_1$, the maximum Kerr phase shift can be written as
\beq\label{deltaPhi}
\Delta\Phi_{\rm NL}(\tilde\omega_1)=\frac{\hbar \omega Q_1 \chi^{(3)}}{4 \epsilon_0 V} g(\tilde\omega_1),
\eeq
where $Q_1=\omega_1/(2|\gamma_1|)$ is the definition of the $Q$-factor of a dipolar resonance in a lossy cavity \cite{kristensen2020modeling}, and
\beq\label{gOmega}
g(\omega)
=\frac{16\omega^6\Delta\omega\varepsilon_2^2(\omega)}{\pi^2}\frac{D_{1m}|I_1|^2}{|\tilde\omega_1(\omega-\tilde\omega_1)|^4}.
\eeq
Using the same parameters as in Fig. \ref{figure1}, we get a maximum phase shift, at resonance, of $\Delta\Phi_{\rm NL}\simeq 3.6\times 10^{-3}$ rad, which is a sizable shift mostly driven by the very small volume of the considered cavity. From this, using Eqs. \eqref{deltaPhi} and \eqref{gOmega}, we can get an estimation for the geometrical factor of $g(\omega_1)\simeq 6\times 10^{-3}$.
The expression in Eq. \eqref{deltaPhi}, which generalized our initial estimate in Eq.~\eqref{eq3}, holds for any general nanocavity in the quantum regime and is the main result of our work.
The resonances are characterised by their $Q$ factors $Q_n$ and the expression $g(\tilde\omega_n)$ is the correction factor that takes into account the geometry of the nanocavity.
\begin{figure}[!t]
\begin{center}
   \includegraphics[width=\linewidth]{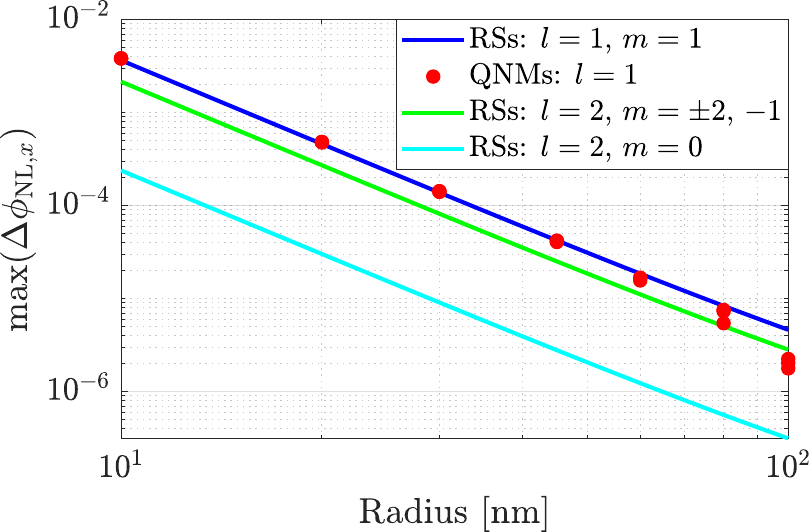}
   \caption{\label{figure2} Comparison of the maximum value of the Kerr nonlinear phase shift $\Delta\Phi_{\rm NL}(\omega)$, as given by Eq. \eqref{deltaPhi}, for both dipolar and quadrupolar RSs as a function of the radius $R$ of the ENZ nanosphere, and for the case, where a single photon is generated in a single RS, corresponding to the dipolar mode with $\ell=1, m=1$ (blue line), a quadrupolar mode with $\ell=2, m=0$ (green line), and $\ell=2, m=\{-1,\pm 2\}$ (cyan line).
   The red dots represent the results obtained from the QNMs analysis for the dipolar modes, each of which has been individually rotated to align along the $x$-axis.
   }
\end{center}
\end{figure}
In our example the $Q$ factor of the ENZ dipolar resonance is quite small, i.e., $Q_1\approx 6$, which only provides a modest amplification of the Kerr phase shift. The results presented in Fig. \ref{figure1} (b), together with Eq. \eqref{deltaPhi}, however, suggest that a careful design of a nanocavity with a large-enough $Q$ factor, attainable, for example, through topological optimisation of dielectric cavities, might provide enough amplification to reach $\Delta\Phi_{\rm NL}\simeq 1$.
With a volume of about $4200\,{\rm nm}^3$, as in this example, in fact, a $Q$-factor of about $10^3$  would suffice to get a Kerr phase shift of the order of $0.5\,{\rm rad}$.

In Fig. \ref{figure2} we present a comparison of the Kerr phase shift $\Delta\Phi_{\rm NL}$ for both the dipolar (blue, solid line) and quadrupolar (green and cyan, solid lines) RS, for different values of the radius $R$ of the ENZ nanosphere.
Again, we set $\mu=\nu=\sigma=\tau=x$, as for Fig. \ref{figure1}.
From the log-log plot in Fig.~\ref{figure2}, a linear trend with a slope of approximately $-3$ is observed, indicating a power-law behavior $\max(\Delta \phi_{\rm NL}) \propto R^{-3}$, in agreement with Eq.~\eqref{deltaPhi}.
Interestingly, despite the fact that the $Q$ factor for the quadrupolar RS is slightly larger than the one of the dipolar mode, i.e., $Q_2=1.04\,Q_1$ and $Q_1=6.67$, we found that $\Delta\phi_{\rm NL}$ for the quadrupole is smaller than the one of the dipole.
This is due to the different values of the geometrical factor $g(\tilde\omega_1)$ for the dipole, vs. $g(\tilde\omega_2)$ for the quadrupole, or, equivalently, from the fact that $\Gamma(\omega_2)<\Gamma(\omega_1)$. To understand this better, we calculate the ratio between the two coupling constants as

\barr
\eta&\equiv&\frac{\Gamma(\omega_2)}{\Gamma(\omega_1)}=\left(\frac{\omega_2}{\omega_1}\right)^8\left[\frac{\varepsilon_2(\omega_2)|I_2|}{\varepsilon_2(\omega_1)|I_1|}\right]^2\left|\frac{\tilde\omega_1}{\tilde\omega_2}\right|^4\left(\frac{D_{2m}}{D_{1m}}\right)\nonumber\\
&\simeq&1.09\left(\frac{D_{2m}}{D_{1m}}\right).
\earr
The quantities $D_{\ell,m}$ come from the integral in Eq. \eqref{eqGamma}, which, for an electric field polarised along the $x$-direction (i.e., for $\mu=\nu=\sigma=\tau=x$) gives
\beq
\int\,d^3r\,|E_{x,\ell,m}(\vett{r})|^4=\frac{D_{\ell,m}}{V}|I_n|^2.
\eeq
These coefficients can be interpreted as the fraction of volume occupied by the $x$-component of the electric field, when projected on the various RSs. For example, the fact that $D_{11}=1$ indicates that the dipolar RS is uniformly extended over the whole sphere, and that $E_x$ aligns perfectly with the dipole mode with $\{\ell=1,m=1\}$. Hence, for the dipole RS the only nonzero coefficient is $D_{11}=1$. For the quadrupolar mode, instead, $D_{20}=5/84$, $D_{21}=D_{22}=D_{2,-2}=15/28$, and $D_{2,-2}=0$, meaning that, in general, the quadrupolar mode occupies a smaller volume in the nanosphere. This is not surprising, since higher order modes of a nanosphere are progressively more confined on its surface. This means, that $\eta\simeq 0.064$ for the $\{\ell=2,m=0\}$ quadrupole, and $\eta\simeq 0.58$ for $\{\ell=2,m=1,\pm 2\}$, and then implies that the Kerr coupling constant for dipolar RSs is the dominant one, since $\Gamma(\omega_2)=\eta\,\Gamma(\omega_1)<\Gamma(\omega_1)$.

\begin{figure*}[t]
\begin{center}
   \includegraphics[width=\textwidth]{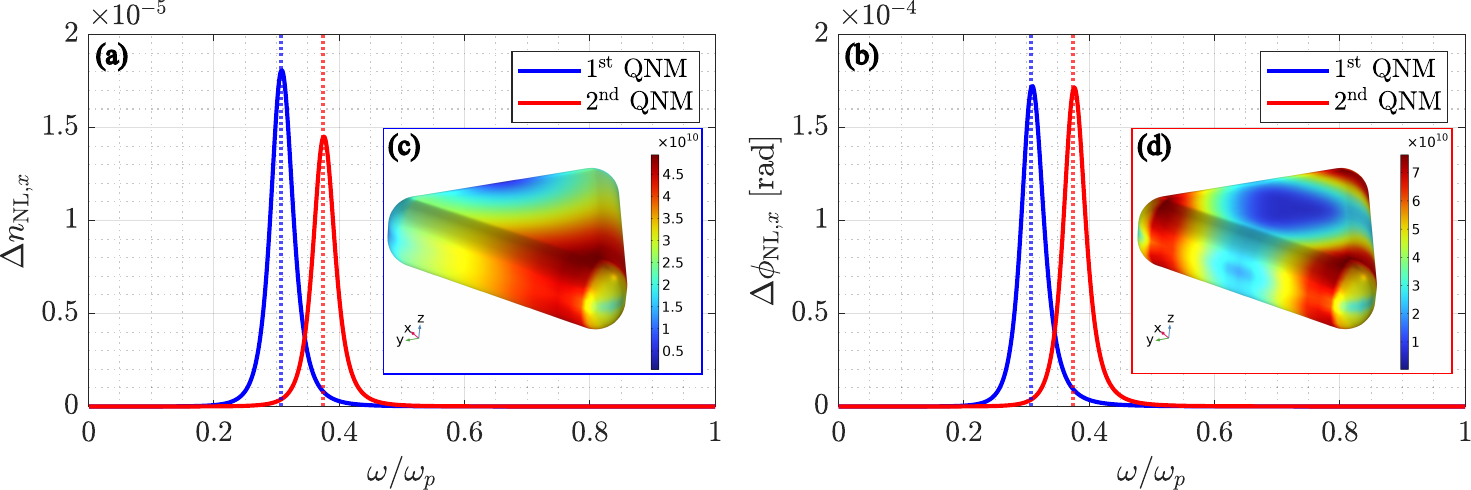}
   \caption{\label{fig:QNMsTriangle}
   (a) Kerr-induced refractive index change $\Delta n_{\rm NL}$ and (b) Kerr-induced phase shift $\Delta\Phi_{\rm NL}$ as a function of the angular frequency of a single photon within the first(second) mode of the triangular cavity oriented along $x$.
   The dotted vertical lines identify the real part of the complex resonant angular frequencies of the first mode shown in (c) and the second mode shown in (d).
   In (c) and (d) is reported the normalized norm of the electric field, $|\tilde{\mathbf{E}}_n| / \sqrt{{\rm QN}_n}$.
   The side length of the equilateral triangle is $70\,{\rm nm}$, its thickness is $20\,{\rm nm}$, and the edge curvature radius is $8\,{\rm nm}$.
   }
\end{center}
\end{figure*}

\section{Kerr shift of ENZ nanocavities of arbitrary shape}\label{sec:QNMs}

In order to validate our theory and extend its prediction to ENZ nanocavities of general (nonspehrical) geometries, we need to compute  the corresponding quasinormal modes (QNMs), which are the fundamental solutions describing the leaky resonances of open electromagnetic systems. Each QNM is associated with a complex eigenfrequency $\tilde{\omega}_n = \omega_n - i\gamma_n$, where the real part $\omega_n$ defines the resonant oscillation frequency, and the imaginary part $\gamma_n$ quantifies the radiative and absorptive losses through the decay rate.
As in the case of RSs, the quality factor can then be directly extracted as $Q = \omega_n/(2|\gamma_n|)$~\cite{wu_nanoscale_2021, kristensen2020modeling, lalanne2019quasinormal}.

A rigorous normalization of the QNMs is essential for accurately computing physically meaningful quantities, such as the modal volume and modal expansion coefficients, and for ensuring the completeness of field representations~\cite{wu_modal_2023, sauvan_normalization_2022}.
However, since QNM fields diverge exponentially in space, numerical evaluation requires the use of Perfectly Matched Layers (PMLs), which transform this divergence into decay within the computational domain.
This makes the fields square-integrable, enabling proper normalization through integration over both the physical region and the PMLs.

In this work, we numerically computed the QNMs of our resonators using the \textsc{MAN} (Modal Analysis of Nanoresonators) toolbox~\cite{wu_modal_2023, wu_qnmnonreciprocal_resonators_2021}, integrated with \textsc{COMSOL Multiphysics} \cite{comsol_sftw}.
The MAN framework provides dedicated solvers (\texttt{QNMEig} and \texttt{QNMPole}) that support dispersive materials modeled via Lorentz–Drude functions, making it suitable for nanostructures with material losses and frequency dispersion. This approach is robust even for low-$Q$ systems, where perturbative techniques often fail.
The QNMs were normalized using the generalized expression~\cite{wu_modal_2023}:
\begin{align} \label{eq:QNM_norm}
{\rm QN}_n = &\iiint_{\Omega \cup \Omega_{\text{PML}}}
\Bigg(
\tilde{\mathbf{E}}_n(\mathbf{r}) \cdot \frac{\partial [\omega \varepsilon(\omega)]}{\partial \omega} \tilde{\mathbf{E}}_n(\mathbf{r})+\\
&-
\tilde{\mathbf{H}}_n(\mathbf{r}) \cdot \frac{\partial (\omega \mu/\varepsilon_0)}{\partial \omega} \tilde{\mathbf{H}}_n(\mathbf{r})
\Bigg) d^3r\,,\nonumber
\end{align}
where $\tilde{\mathbf{E}}_n$ and $\tilde{\mathbf{H}}_n$ are the complex electric and magnetic QNM fields.
Here, the integral is performed in the PML layer $\Omega_{\rm PML}$ and in all the physical space $\Omega$.
Note that, in contrast to the equations reported in \cite{wu_modal_2023}, here $\varepsilon$ denotes the relative permittivity, and ${\rm QN}_n$ is divided by $\varepsilon_0$ to ensure a more direct correspondence with the normalization convention typically adopted in RSs formulations, Eq.~\eqref{normalisation}.
The normalized electric field is then given by $\tilde{\mathbf{E}}_n / \sqrt{{\rm QN}_n}$.
For non-dispersive $\mu$, this normalization could be simplified to~\cite{wu_modal_2023, wu_qnmnonreciprocal_resonators_2021}:
\begin{equation} \label{eq:QNM_norm_simplified}
    {\rm QN}_n = \iiint_{\Omega \cup \Omega_{\text{PML}}}
    \Bigg(
    2\tilde{\mathbf{E}}_n(\mathbf{r}) \cdot \frac{\partial [\omega^2 \varepsilon(\omega)]}{\partial (\omega^2)} \tilde{\mathbf{E}}_n(\mathbf{r})
    \Bigg) d^3r\,,\nonumber
\end{equation}
This results in a very similar normalization previously used in this paper for the RSs modes Eq.~\eqref{normalisation}.

It is important to emphasize that QNM analysis, being inherently numerical, can be applied to complex and asymmetric structures as well. This makes the approach presented in this work broadly applicable and extends its relevance beyond systems with idealized shapes.

\begin{figure*}[!htb]
\begin{center}
   \includegraphics[width=\textwidth]{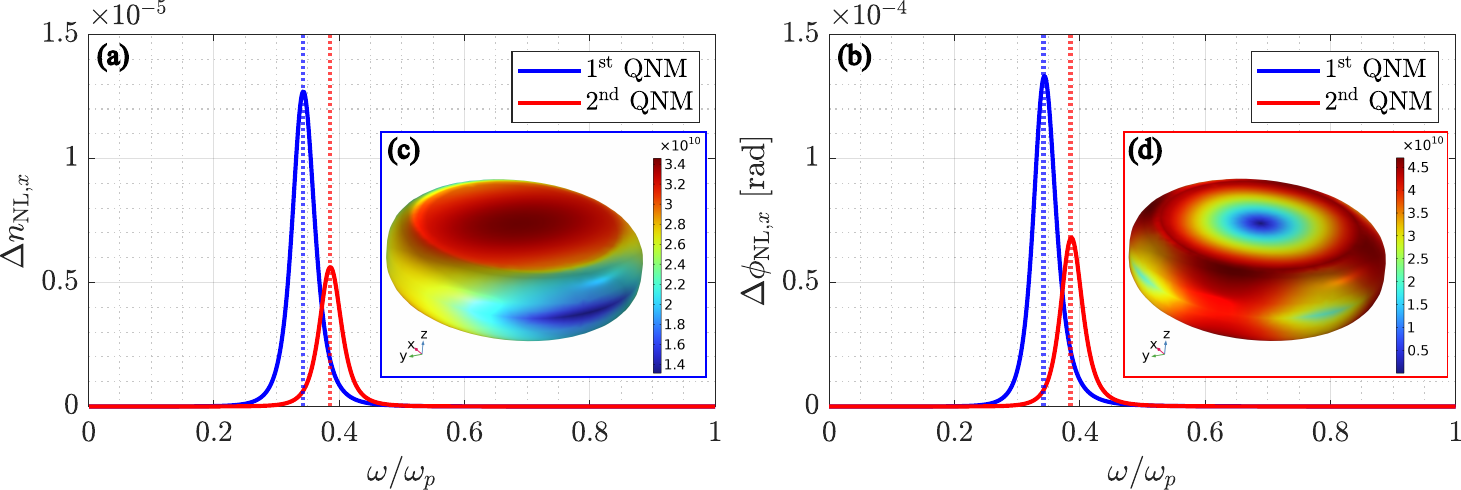}
   \caption{\label{fig:QNMsDisk}
   (a) Kerr-induced refractive index change $\Delta n_{\rm NL}$ and (b) Kerr-induced phase shift $\Delta\Phi_{\rm NL}$ as a function of the angular frequency of a single photon within the first(second) mode of the nanodisk cavity oriented along $x$.
   The dotted vertical lines identify the real part of the complex resonant angular frequencies of the first mode shown in (c) and the second mode shown in (d).
   In (c) and (d) is reported the normalized norm of the electric field, $|\tilde{\mathbf{E}}_n| / \sqrt{{\rm QN}_n}$.
   The diameter of the nanodisk is $70\,{\rm nm}$, its thickness is $20\,{\rm nm}$, and the edge curvature radius is $8\,{\rm nm}$.
   }
\end{center}
\end{figure*}

The modal volume, a key figure of merit for light–matter interaction, is computed as:
\begin{equation} \label{eq:mode_volume}
    \tilde{V}_n(\mathbf{r}, \mathbf{u}) =
    \left[
    2 \left( \tilde{\mathbf{E}}_n \cdot \mathbf{u} \right)^2 / {\rm QN}_n
    \right]^{-1},
\end{equation}
with $\mathbf{u}$ denoting the polarization direction (taken as $\hat{x}$ in our case).
It is worth noticing that $\tilde{V}_n$ is a complex value and depends on position and polarization.

To demonstrate the validity and accuracy of our approach we use here QNM analysis instead of the analytical RSs and compute numerically the Kerr-induced refractive index change and the Kerr-induced phase shift of the dipole mode of the sphere from the general Eq.~\eqref{eqGamma}.
As shown in Figs.~\ref{figure1} and \ref{figure2}, the results obtained with the QNMs closely resemble those obtained with RSs.
However, in Fig.~\ref{figure1}, the peak corresponding to the QNMs is slightly larger and blue-shifted in frequency.
Moreover, Fig.~\ref{figure2} shows that while the two methods are equivalent at small radii, they disagree at larger radii. This apparent discrepancy is due to the asymptotic expansion of the RS formulas for small radii of the sphere as well as to the finite mesh used in the numerical evaluation of Eq.~\eqref{eqGamma}.
Our analysis demonstrate that in the prototypical spherical geometry the results obtained numerically using the QNMs approach are entirely equivalent to the ones obtained analytically based on the RS picture.
This good agreement allows us to apply the general formalism developed in this work, particularly Eq.~\eqref{eqGamma}, to calculate the nonlinear refractive index change and the nonlinear phase shift in more complex structures that can be fabricated on planar substrates using state-of-the-art electron beam lithography.

%We consider in the following the cases of two prototypical nanocavity geometries realized in the ITO dispersive material, that is, the triangular cavity and the nanodisk cavity.
We consider below two prototypical nanocavity geometries fabricated in dispersive ITO material: the triangular cavity and the nanodisk cavity.
In this case, we thus fabricate structures that cannot be solved analytically, and at the same time, we use larger dimensions so that these devices are more easily fabricated.
We chose a side length of $70\,\mathrm{nm}$ for the equilateral triangle equal to the diameter of the nanodisk. Moreover, we set the ITO thickness to $20\,\mathrm{nm}$. Finally, to reduce numerical errors and field divergences caused by the tip effect and, in order to make the structure more realistic, we rounded the edges using a curvature radius of $8\,\mathrm{nm}$.

%The results for the triangular cavity are summarized in Fig.~\ref{fig:QNMsTriangle} (a,b), where we show the nonlinear refractive index change and the phase shift, respectively.
%These values are obtained by using the general equations .....\textcolor{red}{Riccardo please insert the detailed descriptions of your descriptions here.}

Our numerical results for the triangular and the nanodisk cavity are summarized in Figs.~\ref{fig:QNMsTriangle} and \ref{fig:QNMsDisk}.
To reduce the computational resources required to calculate the QNMs of the structures, we exploited the system symmetries by simulating only one quarter of the structure and imposing perfect magnetic boundary conditions to enforce the symmetries (the $xy$ symmetry plane at $z=0$ and the $xz$ symmetry plane at $y=0$).
It is important to note that, when calculating the integrals in Eqs.~\eqref{eqGamma} and \eqref{eq:QNM_norm}, the symmetry factor must be taken into account; using these two symmetry planes, this factor equals 4.

Figure~\ref{fig:QNMsTriangle}(a) shows that the fundamental mode of the triangle exhibits a larger nonlinear refractive index change, $\Delta n_{\rm NL}$, compared to the higher-order mode; however, since the second mode has a larger quality factor $Q$, Fig.~\ref{fig:QNMsTriangle}(b) also indicates that the maximum nonlinear phase shift, $\Delta \phi_{\rm NL}$, is nearly equivalent for both modes.
We also observe that, as expected, the resonance of the higher-order mode occurs at a larger frequency, and these resonances differ from those obtained for the dipole mode of a sphere (Fig.~\ref{figure1}).
This follows from the fact that the plasmonic resonance frequency can be largely tuned by varying the device geometry.

Figure~\ref{fig:QNMsDisk} shows that, for the first two modes of the disk, both $\Delta n_{\rm NL}$ and $\Delta \phi_{\rm NL}$, are greater for the fundamental mode than for the second mode.
In this case, the quality factors of the two modes are more similar compared to the previous case.

Comparing the values of $\Delta \phi_{\rm NL}$ obtained in these two geometries with similar dimensions, it can be observed that although the triangle exhibits a slightly larger nonlinear Kerr shift, the values are very similar and are of the order of $10^{-4}\,{\rm rad}$.
This value is about $20$ times smaller than what obtained for a sphere of $10\,{\rm nm}$ radius; however, if a sphere with $70\,{\rm nm}$ diameter ($R=35\,{\rm nm}$) is considered, which is comparable in size with the two nonspherical geometries analyzed in this section, the Kerr phase shift would be slightly less than $10^{-4}\,{\rm rad}$. Our analysis then demonstrates that the specific shape of the considered particle does not drastically affect the obtainable Kerr phase shift, but instead the overall size of the particle plays the most significant role in determining its single-photon nonlinear response.

\section{Conclusions}\label{conclusions}

In this paper, we analytically and numerically investigated single-photon optical responses of resonant nanocavities with dispersive ENZ materials and Kerr-type nonlinearity.
In particular, we formulated the cavity problem within the rigorous quantum Langevin-noise approach using the Green's tensor quantization method and obtained closed-form analytical solutions for the nonlinear phase shifts of sub-wavelength spherical cavities at the single-photon level.
We also extended our work to the case of non-spherical cavity geometries based on the numerical calculation of quasi-normal modes. 
Our findings demonstrate a Kerr phase shift $\Delta\Phi_{\rm NL}\simeq 3.6\times 10^{-3}$ rad in ITO spherical nanocavities with $10\,{\rm nm}$ radius, and $\Delta\Phi_{\rm NL}\approx 10^{-4}$ rad
in triangular and disk cavity geometries with the side length of the triangle or the diameter of the disk equal to $70\,{\rm nm}$.
Moreover, our analysis indicates that a careful design of a nonlinear nanocavity with a volume of about $4200\,{\rm nm}^3$ and a $Q\approx 10^{3}$, attainable, for example, through topological optimisation of dielectric cavities, might provide enough amplification to reach a nonlinear Kerr phase shift of the order of $0.5\,{\rm rad}$ at the single photon level.
Furthermore, although this work focuses on the ITO as a candidate ENZ material with large Kerr-type nonlinear response, the approach that we developed here is general and can be applied to additional ENZ materials with Drude-Sommerfeld causal dispersion, including cadmium oxide (CdO), which has recently attracted significant interest due to its reduced optical losses and large optical nonlinearity \cite{cleri2021mid, schrecengost2024large}.
Finally, our theoretical analysis provides a robust and accurate methodology for future studies of nonlinear quantum electrodynamics in realistic ENZ devices and nanostructures that are important to emerging quantum technology applications, including on-chip single-photon nondemolition detection, quantum sensing, controlled quantum gates, as well as for the engineering of photon blockade effects at the nanoscale.

\begin{acknowledgments}
L.D.N. acknowledges the support from the U.S. Army Research Office under award number W911NF2210110.
 M. O. acknowledges the financial support from the Research Council of Finland Flagship Programme (PREIN - decision Grant No. 320165).
\end{acknowledgments}

\appendix

\section*{Appendix A: Derivation of Eq. (\ref{eq57})}\label{appendix1}

To derive Eq. \eqref{eq57}, let us first notice that $E_{n,\alpha}(\vett{r})E_{q,\alpha}(\vett{r})=\vett{E}_n(\vett{r})\cdot\vett{E}_q(\vett{r})$, since summation over repeated indices is implicitly understood. Since the RS are a complete basis, they are mutually orthogonal. This allows us to write
\beq
\int_V\,d^3r\,\vett{E}_n(\vett{r})\cdot\vett{E}_q(\vett{r})=\delta_{nq}I,
\eeq
where $I$ is the integral above with $n=q$, which is what we calculate in this Appendix. Using Eq. \eqref{eq52} we can then write the volume integral as
\barr\label{eqAppC1}
I&=&\int_V\,d^3r\,\vett{E}_n(\vett{r})\cdot\vett{E}_n(\vett{r})=\frac{(A^{\rm TM}_{\ell,n})^2}{k_nk_q}\Bigg[\ell^2(\ell+1)^2I_r^{(1)}I_{\theta,\varphi}^{(1)}\nonumber\\
&+&I_r^{(2)}I_{\theta,\varphi}^{(2)}\Bigg],
\earr
where
\bseq
\begin{align}
I_r^{(1)}&=\int_0^R\,dr\,Z_{\ell}(r,\tilde\omega_n)Z_{\ell}(r,\tilde\omega_n),\\
I_r^{(2)}&=\int_0^R\,dr
\,\frac{\partial}{\partial r}\Bigg[r Z_{\ell}(r,\tilde\omega_n)\Bigg]\frac{\partial}{\partial r}\Bigg[r Z_{\ell}(r,\tilde\omega_n)\Bigg],\\
I_{\theta,\varphi}^{(1)}&=\int_0^{2\pi}d\varphi\,\int_0^{\pi}\,d\theta\,\sin\theta\,Y_{\ell}^m(\theta,\varphi)Y_{\ell}^m(\theta,\varphi),\\
I_{\theta,\varphi}^{(2)}&=\int_0^{2\pi}d\varphi\,\int_0^{\pi}\,d\theta\,\sin\theta
\Bigg[\frac{1}{\sin^2\theta}\frac{\partial}{\partial\varphi} Y_{\ell}^m(\theta,\varphi)\frac{\partial}{\partial\varphi} Y_{\ell}^m(\theta,\varphi)\nonumber\\
&+\frac{\partial}{\partial\theta}Y_{\ell}^m(\theta,\varphi)\frac{\partial}{\partial\theta}Y_{\ell}^m(\theta,\varphi)\Bigg].
\end{align}
\eseq
It is not difficult to show, following Ref. \cite{stratton}, that $I_{\theta,\varphi}^{(1)}=1$ and $I_{\theta,\varphi}=\ell(\ell+1)$, and that the radial integral then reduces to $I_r=\ell^2(\ell+1)^2I_r^{(1)}+\ell(\ell+1)I_r^{(2)}$. After a little algebra, and by introducing the quantity $x=k_nR$ (with $k_n\equiv k(\tilde\omega_n)=\tilde\omega_n\sqrt{\varepsilon(\tilde\omega_n)}/c$) the radial integral can be written in the following, compact form
\beq
I_r=\frac{\ell(\ell+1)R^3}{x^2}I_{\rho}(x),
\eeq
where
\barr
I_{\rho}(x)&=&\Bigg[(\ell+1)-\frac{x j_{\ell+1}(x)}{j_{\ell}(x)}\nonumber\\
&-&\frac{x^2}{2}\frac{j_{\ell-1}(x)j_{\ell+2}(x)}{j_{\ell}^2(x)}+\frac{x^2}{2}\Bigg].
\earr
If we then introduce the quantities
\bseq
\begin{align}
\xi_{\ell}(x)&=\frac{I_{\rho}}{x^2},\\
s_0(\tilde\omega_n)&=\varepsilon(\tilde\omega_n)D_{\ell}(\tilde\omega_n)[\varepsilon(\tilde\omega_n)-1],
\end{align}
\eseq
we can write the integral as in Eq. \eqref{eq57}, if we Taylor expand the quantity $I_{\rho}(x)$ in the definition of $\xi_{\ell}(x)$, and $D_{\ell}(\tilde\omega_n)$ in the expression of $s_0(\tilde\omega_n)$ to the lowest relevant order in $x\ll 1$, obtaining
\bseq\label{eq83}
\begin{align}
    \xi_{\ell}(x)&\simeq\frac{(\ell+1)}{x^2},\\
    s_0(\tilde\omega_n)&\simeq \frac{\ell+1}{x^2}h(\tilde\omega_n),
\end{align}
\eseq
where $h(\omega)=[\varepsilon(\omega)-1][1+\ell+\ell\varepsilon(\tilde\omega)]+\omega\varepsilon'(\omega)$, and Eq. \eqref{eq57} can be written simply as
\barr
I=\frac{\xi_{\ell}(x)}{s_0(\tilde\omega_n)}&\simeq&\frac{1}{h(\tilde\omega_n)}=\frac{1}{\tilde\omega_n\varepsilon'(\tilde\omega_n)}.
\earr

\bibliography{references} %COMMENT before submitting

\begin{thebibliography}{72}%
\makeatletter
\providecommand \@ifxundefined [1]{%
 \@ifx{#1\undefined}
}%
\providecommand \@ifnum [1]{%
 \ifnum #1\expandafter \@firstoftwo
 \else \expandafter \@secondoftwo
 \fi
}%
\providecommand \@ifx [1]{%
 \ifx #1\expandafter \@firstoftwo
 \else \expandafter \@secondoftwo
 \fi
}%
\providecommand \natexlab [1]{#1}%
\providecommand \enquote  [1]{``#1''}%
\providecommand \bibnamefont  [1]{#1}%
\providecommand \bibfnamefont [1]{#1}%
\providecommand \citenamefont [1]{#1}%
\providecommand \href@noop [0]{\@secondoftwo}%
\providecommand \href [0]{\begingroup \@sanitize@url \@href}%
\providecommand \@href[1]{\@@startlink{#1}\@@href}%
\providecommand \@@href[1]{\endgroup#1\@@endlink}%
\providecommand \@sanitize@url [0]{\catcode `\\12\catcode `\$12\catcode `\&12\catcode `\#12\catcode `\^12\catcode `\_12\catcode `\%12\relax}%
\providecommand \@@startlink[1]{}%
\providecommand \@@endlink[0]{}%
\providecommand \url  [0]{\begingroup\@sanitize@url \@url }%
\providecommand \@url [1]{\endgroup\@href {#1}{\urlprefix }}%
\providecommand \urlprefix  [0]{URL }%
\providecommand \Eprint [0]{\href }%
\providecommand \doibase [0]{https://doi.org/}%
\providecommand \selectlanguage [0]{\@gobble}%
\providecommand \bibinfo  [0]{\@secondoftwo}%
\providecommand \bibfield  [0]{\@secondoftwo}%
\providecommand \translation [1]{[#1]}%
\providecommand \BibitemOpen [0]{}%
\providecommand \bibitemStop [0]{}%
\providecommand \bibitemNoStop [0]{.\EOS\space}%
\providecommand \EOS [0]{\spacefactor3000\relax}%
\providecommand \BibitemShut  [1]{\csname bibitem#1\endcsname}%
\let\auto@bib@innerbib\@empty
%</preamble>
\bibitem [{\citenamefont {O'brien}\ \emph {et~al.}(2009)\citenamefont {O'brien}, \citenamefont {Furusawa},\ and\ \citenamefont {Vu{\v{c}}kovi{\'c}}}]{o2009photonic}%
  \BibitemOpen
  \bibfield  {author} {\bibinfo {author} {\bibfnamefont {J.~L.}\ \bibnamefont {O'brien}}, \bibinfo {author} {\bibfnamefont {A.}~\bibnamefont {Furusawa}},\ and\ \bibinfo {author} {\bibfnamefont {J.}~\bibnamefont {Vu{\v{c}}kovi{\'c}}},\ }\bibfield  {title} {\bibinfo {title} {Photonic quantum technologies},\ }\href@noop {} {\bibfield  {journal} {\bibinfo  {journal} {Nature photonics}\ }\textbf {\bibinfo {volume} {3}},\ \bibinfo {pages} {687} (\bibinfo {year} {2009})}\BibitemShut {NoStop}%
\bibitem [{\citenamefont {Dousse}\ \emph {et~al.}(2010)\citenamefont {Dousse}, \citenamefont {Suffczy{\'n}ski}, \citenamefont {Beveratos}, \citenamefont {Krebs}, \citenamefont {Lema{\^\i}tre}, \citenamefont {Sagnes}, \citenamefont {Bloch}, \citenamefont {Voisin},\ and\ \citenamefont {Senellart}}]{dousse2010ultrabright}%
  \BibitemOpen
  \bibfield  {author} {\bibinfo {author} {\bibfnamefont {A.}~\bibnamefont {Dousse}}, \bibinfo {author} {\bibfnamefont {J.}~\bibnamefont {Suffczy{\'n}ski}}, \bibinfo {author} {\bibfnamefont {A.}~\bibnamefont {Beveratos}}, \bibinfo {author} {\bibfnamefont {O.}~\bibnamefont {Krebs}}, \bibinfo {author} {\bibfnamefont {A.}~\bibnamefont {Lema{\^\i}tre}}, \bibinfo {author} {\bibfnamefont {I.}~\bibnamefont {Sagnes}}, \bibinfo {author} {\bibfnamefont {J.}~\bibnamefont {Bloch}}, \bibinfo {author} {\bibfnamefont {P.}~\bibnamefont {Voisin}},\ and\ \bibinfo {author} {\bibfnamefont {P.}~\bibnamefont {Senellart}},\ }\bibfield  {title} {\bibinfo {title} {Ultrabright source of entangled photon pairs},\ }\href@noop {} {\bibfield  {journal} {\bibinfo  {journal} {Nature}\ }\textbf {\bibinfo {volume} {466}},\ \bibinfo {pages} {217} (\bibinfo {year} {2010})}\BibitemShut {NoStop}%
\bibitem [{\citenamefont {Tonndorf}\ \emph {et~al.}(2017)\citenamefont {Tonndorf}, \citenamefont {Del Pozo-Zamudio}, \citenamefont {Gruhler}, \citenamefont {Kern}, \citenamefont {Schmidt}, \citenamefont {Dmitriev}, \citenamefont {Bakhtinov}, \citenamefont {Tartakovskii}, \citenamefont {Pernice}, \citenamefont {Michaelis~de Vasconcellos} \emph {et~al.}}]{tonndorf2017chip}%
  \BibitemOpen
  \bibfield  {author} {\bibinfo {author} {\bibfnamefont {P.}~\bibnamefont {Tonndorf}}, \bibinfo {author} {\bibfnamefont {O.}~\bibnamefont {Del Pozo-Zamudio}}, \bibinfo {author} {\bibfnamefont {N.}~\bibnamefont {Gruhler}}, \bibinfo {author} {\bibfnamefont {J.}~\bibnamefont {Kern}}, \bibinfo {author} {\bibfnamefont {R.}~\bibnamefont {Schmidt}}, \bibinfo {author} {\bibfnamefont {A.~I.}\ \bibnamefont {Dmitriev}}, \bibinfo {author} {\bibfnamefont {A.~P.}\ \bibnamefont {Bakhtinov}}, \bibinfo {author} {\bibfnamefont {A.~I.}\ \bibnamefont {Tartakovskii}}, \bibinfo {author} {\bibfnamefont {W.}~\bibnamefont {Pernice}}, \bibinfo {author} {\bibfnamefont {S.}~\bibnamefont {Michaelis~de Vasconcellos}}, \emph {et~al.},\ }\bibfield  {title} {\bibinfo {title} {On-chip waveguide coupling of a layered semiconductor single-photon source},\ }\href@noop {} {\bibfield  {journal} {\bibinfo  {journal} {Nano letters}\ }\textbf {\bibinfo {volume} {17}},\ \bibinfo {pages} {5446} (\bibinfo {year} {2017})}\BibitemShut {NoStop}%
\bibitem [{\citenamefont {Eisaman}\ \emph {et~al.}(2011)\citenamefont {Eisaman}, \citenamefont {Fan}, \citenamefont {Migdall},\ and\ \citenamefont {Polyakov}}]{eisaman2011invited}%
  \BibitemOpen
  \bibfield  {author} {\bibinfo {author} {\bibfnamefont {M.~D.}\ \bibnamefont {Eisaman}}, \bibinfo {author} {\bibfnamefont {J.}~\bibnamefont {Fan}}, \bibinfo {author} {\bibfnamefont {A.}~\bibnamefont {Migdall}},\ and\ \bibinfo {author} {\bibfnamefont {S.~V.}\ \bibnamefont {Polyakov}},\ }\bibfield  {title} {\bibinfo {title} {Invited review article: Single-photon sources and detectors},\ }\href@noop {} {\bibfield  {journal} {\bibinfo  {journal} {Review of scientific instruments}\ }\textbf {\bibinfo {volume} {82}},\ \bibinfo {pages} {071101} (\bibinfo {year} {2011})}\BibitemShut {NoStop}%
\bibitem [{\citenamefont {Boyer}\ \emph {et~al.}(2008)\citenamefont {Boyer}, \citenamefont {Marino}, \citenamefont {Pooser},\ and\ \citenamefont {Lett}}]{boyer2008entangled}%
  \BibitemOpen
  \bibfield  {author} {\bibinfo {author} {\bibfnamefont {V.}~\bibnamefont {Boyer}}, \bibinfo {author} {\bibfnamefont {A.~M.}\ \bibnamefont {Marino}}, \bibinfo {author} {\bibfnamefont {R.~C.}\ \bibnamefont {Pooser}},\ and\ \bibinfo {author} {\bibfnamefont {P.~D.}\ \bibnamefont {Lett}},\ }\bibfield  {title} {\bibinfo {title} {Entangled images from four-wave mixing},\ }\href@noop {} {\bibfield  {journal} {\bibinfo  {journal} {Science}\ }\textbf {\bibinfo {volume} {321}},\ \bibinfo {pages} {544} (\bibinfo {year} {2008})}\BibitemShut {NoStop}%
\bibitem [{\citenamefont {Zhang}\ \emph {et~al.}(2022)\citenamefont {Zhang}, \citenamefont {Ma}, \citenamefont {Parry}, \citenamefont {Cai}, \citenamefont {Camacho-Morales}, \citenamefont {Xu}, \citenamefont {Neshev},\ and\ \citenamefont {Sukhorukov}}]{zhang2022spatially}%
  \BibitemOpen
  \bibfield  {author} {\bibinfo {author} {\bibfnamefont {J.}~\bibnamefont {Zhang}}, \bibinfo {author} {\bibfnamefont {J.}~\bibnamefont {Ma}}, \bibinfo {author} {\bibfnamefont {M.}~\bibnamefont {Parry}}, \bibinfo {author} {\bibfnamefont {M.}~\bibnamefont {Cai}}, \bibinfo {author} {\bibfnamefont {R.}~\bibnamefont {Camacho-Morales}}, \bibinfo {author} {\bibfnamefont {L.}~\bibnamefont {Xu}}, \bibinfo {author} {\bibfnamefont {D.~N.}\ \bibnamefont {Neshev}},\ and\ \bibinfo {author} {\bibfnamefont {A.~A.}\ \bibnamefont {Sukhorukov}},\ }\bibfield  {title} {\bibinfo {title} {Spatially entangled photon pairs from lithium niobate nonlocal metasurfaces},\ }\href@noop {} {\bibfield  {journal} {\bibinfo  {journal} {Science Advances}\ }\textbf {\bibinfo {volume} {8}},\ \bibinfo {pages} {eabq4240} (\bibinfo {year} {2022})}\BibitemShut {NoStop}%
\bibitem [{\citenamefont {Munro}\ \emph {et~al.}(2005)\citenamefont {Munro}, \citenamefont {Nemoto}, \citenamefont {Beausoleil},\ and\ \citenamefont {Spiller}}]{munro2005high}%
  \BibitemOpen
  \bibfield  {author} {\bibinfo {author} {\bibfnamefont {W.}~\bibnamefont {Munro}}, \bibinfo {author} {\bibfnamefont {K.}~\bibnamefont {Nemoto}}, \bibinfo {author} {\bibfnamefont {R.}~\bibnamefont {Beausoleil}},\ and\ \bibinfo {author} {\bibfnamefont {T.}~\bibnamefont {Spiller}},\ }\bibfield  {title} {\bibinfo {title} {High-efficiency quantum-nondemolition single-photon-number-resolving detector},\ }\href@noop {} {\bibfield  {journal} {\bibinfo  {journal} {Physical Review A}\ }\textbf {\bibinfo {volume} {71}},\ \bibinfo {pages} {033819} (\bibinfo {year} {2005})}\BibitemShut {NoStop}%
\bibitem [{\citenamefont {Flayac}\ \emph {et~al.}(2015)\citenamefont {Flayac}, \citenamefont {Gerace},\ and\ \citenamefont {Savona}}]{flayac2015all}%
  \BibitemOpen
  \bibfield  {author} {\bibinfo {author} {\bibfnamefont {H.}~\bibnamefont {Flayac}}, \bibinfo {author} {\bibfnamefont {D.}~\bibnamefont {Gerace}},\ and\ \bibinfo {author} {\bibfnamefont {V.}~\bibnamefont {Savona}},\ }\bibfield  {title} {\bibinfo {title} {An all-silicon single-photon source by unconventional photon blockade},\ }\href@noop {} {\bibfield  {journal} {\bibinfo  {journal} {Scientific reports}\ }\textbf {\bibinfo {volume} {5}},\ \bibinfo {pages} {1} (\bibinfo {year} {2015})}\BibitemShut {NoStop}%
\bibitem [{\citenamefont {Ferretti}\ and\ \citenamefont {Gerace}(2012)}]{ferretti2012single}%
  \BibitemOpen
  \bibfield  {author} {\bibinfo {author} {\bibfnamefont {S.}~\bibnamefont {Ferretti}}\ and\ \bibinfo {author} {\bibfnamefont {D.}~\bibnamefont {Gerace}},\ }\bibfield  {title} {\bibinfo {title} {Single-photon nonlinear optics with kerr-type nanostructured materials},\ }\href@noop {} {\bibfield  {journal} {\bibinfo  {journal} {Physical Review B}\ }\textbf {\bibinfo {volume} {85}},\ \bibinfo {pages} {033303} (\bibinfo {year} {2012})}\BibitemShut {NoStop}%
\bibitem [{\citenamefont {Kok}\ \emph {et~al.}(2002)\citenamefont {Kok}, \citenamefont {Lee},\ and\ \citenamefont {Dowling}}]{PhysRevA.66.063814}%
  \BibitemOpen
  \bibfield  {author} {\bibinfo {author} {\bibfnamefont {P.}~\bibnamefont {Kok}}, \bibinfo {author} {\bibfnamefont {H.}~\bibnamefont {Lee}},\ and\ \bibinfo {author} {\bibfnamefont {J.~P.}\ \bibnamefont {Dowling}},\ }\bibfield  {title} {\bibinfo {title} {Single-photon quantum-nondemolition detectors constructed with linear optics and projective measurements},\ }\href {https://doi.org/10.1103/PhysRevA.66.063814} {\bibfield  {journal} {\bibinfo  {journal} {Phys. Rev. A}\ }\textbf {\bibinfo {volume} {66}},\ \bibinfo {pages} {063814} (\bibinfo {year} {2002})}\BibitemShut {NoStop}%
\bibitem [{\citenamefont {Imoto}\ \emph {et~al.}(1985)\citenamefont {Imoto}, \citenamefont {Haus},\ and\ \citenamefont {Yamamoto}}]{imoto1985quantum}%
  \BibitemOpen
  \bibfield  {author} {\bibinfo {author} {\bibfnamefont {N.}~\bibnamefont {Imoto}}, \bibinfo {author} {\bibfnamefont {H.}~\bibnamefont {Haus}},\ and\ \bibinfo {author} {\bibfnamefont {Y.}~\bibnamefont {Yamamoto}},\ }\bibfield  {title} {\bibinfo {title} {Quantum nondemolition measurement of the photon number via the optical kerr effect},\ }\href@noop {} {\bibfield  {journal} {\bibinfo  {journal} {Physical Review A}\ }\textbf {\bibinfo {volume} {32}},\ \bibinfo {pages} {2287} (\bibinfo {year} {1985})}\BibitemShut {NoStop}%
\bibitem [{\citenamefont {Xiao}\ \emph {et~al.}(2008)\citenamefont {Xiao}, \citenamefont {{\"O}zdemir}, \citenamefont {Gaddam}, \citenamefont {Dong}, \citenamefont {Imoto},\ and\ \citenamefont {Yang}}]{xiao2008quantum}%
  \BibitemOpen
  \bibfield  {author} {\bibinfo {author} {\bibfnamefont {Y.-F.}\ \bibnamefont {Xiao}}, \bibinfo {author} {\bibfnamefont {{\c{S}}.~K.}\ \bibnamefont {{\"O}zdemir}}, \bibinfo {author} {\bibfnamefont {V.}~\bibnamefont {Gaddam}}, \bibinfo {author} {\bibfnamefont {C.-H.}\ \bibnamefont {Dong}}, \bibinfo {author} {\bibfnamefont {N.}~\bibnamefont {Imoto}},\ and\ \bibinfo {author} {\bibfnamefont {L.}~\bibnamefont {Yang}},\ }\bibfield  {title} {\bibinfo {title} {Quantum nondemolition measurement of photon number via optical kerr effect in an ultra-high-q microtoroid cavity},\ }\href@noop {} {\bibfield  {journal} {\bibinfo  {journal} {Optics Express}\ }\textbf {\bibinfo {volume} {16}},\ \bibinfo {pages} {21462} (\bibinfo {year} {2008})}\BibitemShut {NoStop}%
\bibitem [{\citenamefont {Braginsky}\ and\ \citenamefont {Khalili}(1996)}]{braginsky1996quantum}%
  \BibitemOpen
  \bibfield  {author} {\bibinfo {author} {\bibfnamefont {V.~B.}\ \bibnamefont {Braginsky}}\ and\ \bibinfo {author} {\bibfnamefont {F.~Y.}\ \bibnamefont {Khalili}},\ }\bibfield  {title} {\bibinfo {title} {Quantum nondemolition measurements: the route from toys to tools},\ }\href@noop {} {\bibfield  {journal} {\bibinfo  {journal} {Reviews of Modern Physics}\ }\textbf {\bibinfo {volume} {68}},\ \bibinfo {pages} {1} (\bibinfo {year} {1996})}\BibitemShut {NoStop}%
\bibitem [{\citenamefont {Huttner}\ and\ \citenamefont {Barnett}(1992)}]{huttner1992quantization}%
  \BibitemOpen
  \bibfield  {author} {\bibinfo {author} {\bibfnamefont {B.}~\bibnamefont {Huttner}}\ and\ \bibinfo {author} {\bibfnamefont {S.~M.}\ \bibnamefont {Barnett}},\ }\bibfield  {title} {\bibinfo {title} {Quantization of the electromagnetic field in dielectrics},\ }\href@noop {} {\bibfield  {journal} {\bibinfo  {journal} {Physical Review A}\ }\textbf {\bibinfo {volume} {46}},\ \bibinfo {pages} {4306} (\bibinfo {year} {1992})}\BibitemShut {NoStop}%
\bibitem [{\citenamefont {Hopfield}(1958)}]{hopfield1958theory}%
  \BibitemOpen
  \bibfield  {author} {\bibinfo {author} {\bibfnamefont {J.}~\bibnamefont {Hopfield}},\ }\bibfield  {title} {\bibinfo {title} {Theory of the contribution of excitons to the complex dielectric constant of crystals},\ }\href@noop {} {\bibfield  {journal} {\bibinfo  {journal} {Physical Review}\ }\textbf {\bibinfo {volume} {112}},\ \bibinfo {pages} {1555} (\bibinfo {year} {1958})}\BibitemShut {NoStop}%
\bibitem [{\citenamefont {Knoll}\ \emph {et~al.}(2000)\citenamefont {Knoll}, \citenamefont {Scheel},\ and\ \citenamefont {Welsch}}]{knoll2000qed}%
  \BibitemOpen
  \bibfield  {author} {\bibinfo {author} {\bibfnamefont {L.}~\bibnamefont {Knoll}}, \bibinfo {author} {\bibfnamefont {S.}~\bibnamefont {Scheel}},\ and\ \bibinfo {author} {\bibfnamefont {D.-G.}\ \bibnamefont {Welsch}},\ }\bibfield  {title} {\bibinfo {title} {Qed in dispersing and absorbing media},\ }\href@noop {} {\bibfield  {journal} {\bibinfo  {journal} {arXiv preprint quant-ph/0006121}\ } (\bibinfo {year} {2000})}\BibitemShut {NoStop}%
\bibitem [{\citenamefont {Bechler}(2006)}]{bechler2006path}%
  \BibitemOpen
  \bibfield  {author} {\bibinfo {author} {\bibfnamefont {A.}~\bibnamefont {Bechler}},\ }\bibfield  {title} {\bibinfo {title} {Path-integral quantization of the electromagnetic field in the hopfield dielectric beyond dipole approximation},\ }\href@noop {} {\bibfield  {journal} {\bibinfo  {journal} {Journal of Physics A: Mathematical and General}\ }\textbf {\bibinfo {volume} {39}},\ \bibinfo {pages} {13553} (\bibinfo {year} {2006})}\BibitemShut {NoStop}%
\bibitem [{\citenamefont {Vogel}\ and\ \citenamefont {Welsch}(2006)}]{vogel2006quantum}%
  \BibitemOpen
  \bibfield  {author} {\bibinfo {author} {\bibfnamefont {W.}~\bibnamefont {Vogel}}\ and\ \bibinfo {author} {\bibfnamefont {D.-G.}\ \bibnamefont {Welsch}},\ }\href@noop {} {\emph {\bibinfo {title} {Quantum optics}}}\ (\bibinfo  {publisher} {John Wiley \& Sons},\ \bibinfo {year} {2006})\BibitemShut {NoStop}%
\bibitem [{\citenamefont {Dung}\ \emph {et~al.}(2006)\citenamefont {Dung}, \citenamefont {Buhmann},\ and\ \citenamefont {Welsch}}]{dung2006}%
  \BibitemOpen
  \bibfield  {author} {\bibinfo {author} {\bibfnamefont {H.~T.}\ \bibnamefont {Dung}}, \bibinfo {author} {\bibfnamefont {S.~Y.}\ \bibnamefont {Buhmann}},\ and\ \bibinfo {author} {\bibfnamefont {D.-G.}\ \bibnamefont {Welsch}},\ }\bibfield  {title} {\bibinfo {title} {Local-field correction to the spontaneous decay rate of atoms embedded in bodies of finite size},\ }\href@noop {} {\bibfield  {journal} {\bibinfo  {journal} {Phys. Rev. A}\ }\textbf {\bibinfo {volume} {74}},\ \bibinfo {pages} {023803} (\bibinfo {year} {2006})}\BibitemShut {NoStop}%
\bibitem [{\citenamefont {Gruner}\ and\ \citenamefont {Welsch}(1996)}]{gruner1996green}%
  \BibitemOpen
  \bibfield  {author} {\bibinfo {author} {\bibfnamefont {T.}~\bibnamefont {Gruner}}\ and\ \bibinfo {author} {\bibfnamefont {D.-G.}\ \bibnamefont {Welsch}},\ }\bibfield  {title} {\bibinfo {title} {Green-function approach to the radiation-field quantization for homogeneous and inhomogeneous kramers-kronig dielectrics},\ }\href@noop {} {\bibfield  {journal} {\bibinfo  {journal} {Physical Review A}\ }\textbf {\bibinfo {volume} {53}},\ \bibinfo {pages} {1818} (\bibinfo {year} {1996})}\BibitemShut {NoStop}%
\bibitem [{\citenamefont {Difallah}\ \emph {et~al.}(2019)\citenamefont {Difallah}, \citenamefont {Szameit},\ and\ \citenamefont {Ornigotti}}]{difallah2019path}%
  \BibitemOpen
  \bibfield  {author} {\bibinfo {author} {\bibfnamefont {M.}~\bibnamefont {Difallah}}, \bibinfo {author} {\bibfnamefont {A.}~\bibnamefont {Szameit}},\ and\ \bibinfo {author} {\bibfnamefont {M.}~\bibnamefont {Ornigotti}},\ }\bibfield  {title} {\bibinfo {title} {Path-integral description of quantum nonlinear optics in arbitrary media},\ }\href@noop {} {\bibfield  {journal} {\bibinfo  {journal} {Physical Review A}\ }\textbf {\bibinfo {volume} {100}},\ \bibinfo {pages} {053845} (\bibinfo {year} {2019})}\BibitemShut {NoStop}%
\bibitem [{\citenamefont {Crosse}\ and\ \citenamefont {Scheel}(2010)}]{crosse2010effective}%
  \BibitemOpen
  \bibfield  {author} {\bibinfo {author} {\bibfnamefont {J.}~\bibnamefont {Crosse}}\ and\ \bibinfo {author} {\bibfnamefont {S.}~\bibnamefont {Scheel}},\ }\bibfield  {title} {\bibinfo {title} {Effective nonlinear hamiltonians in dielectric media},\ }\href@noop {} {\bibfield  {journal} {\bibinfo  {journal} {Physical Review A—Atomic, Molecular, and Optical Physics}\ }\textbf {\bibinfo {volume} {81}},\ \bibinfo {pages} {033815} (\bibinfo {year} {2010})}\BibitemShut {NoStop}%
\bibitem [{\citenamefont {Scheel}\ and\ \citenamefont {Welsch}(2006{\natexlab{a}})}]{scheel2006causal}%
  \BibitemOpen
  \bibfield  {author} {\bibinfo {author} {\bibfnamefont {S.}~\bibnamefont {Scheel}}\ and\ \bibinfo {author} {\bibfnamefont {D.-G.}\ \bibnamefont {Welsch}},\ }\bibfield  {title} {\bibinfo {title} {Causal nonlinear quantum optics},\ }\href@noop {} {\bibfield  {journal} {\bibinfo  {journal} {Journal of Physics B: Atomic, Molecular and Optical Physics}\ }\textbf {\bibinfo {volume} {39}},\ \bibinfo {pages} {S711} (\bibinfo {year} {2006}{\natexlab{a}})}\BibitemShut {NoStop}%
\bibitem [{\citenamefont {Scheel}\ and\ \citenamefont {Welsch}(2006{\natexlab{b}})}]{scheel2006quantum}%
  \BibitemOpen
  \bibfield  {author} {\bibinfo {author} {\bibfnamefont {S.}~\bibnamefont {Scheel}}\ and\ \bibinfo {author} {\bibfnamefont {D.-G.}\ \bibnamefont {Welsch}},\ }\bibfield  {title} {\bibinfo {title} {Quantum theory of light and noise polarization in nonlinear optics},\ }\href@noop {} {\bibfield  {journal} {\bibinfo  {journal} {Physical Review Letters}\ }\textbf {\bibinfo {volume} {96}},\ \bibinfo {pages} {073601} (\bibinfo {year} {2006}{\natexlab{b}})}\BibitemShut {NoStop}%
\bibitem [{\citenamefont {Krsti{\'c}}\ \emph {et~al.}(2023)\citenamefont {Krsti{\'c}}, \citenamefont {Setzpfandt},\ and\ \citenamefont {Saravi}}]{krstic2023nonperturbative}%
  \BibitemOpen
  \bibfield  {author} {\bibinfo {author} {\bibfnamefont {A.}~\bibnamefont {Krsti{\'c}}}, \bibinfo {author} {\bibfnamefont {F.}~\bibnamefont {Setzpfandt}},\ and\ \bibinfo {author} {\bibfnamefont {S.}~\bibnamefont {Saravi}},\ }\bibfield  {title} {\bibinfo {title} {Nonperturbative theory of spontaneous parametric down-conversion in open and dispersive optical systems},\ }\href@noop {} {\bibfield  {journal} {\bibinfo  {journal} {Physical Review Research}\ }\textbf {\bibinfo {volume} {5}},\ \bibinfo {pages} {043228} (\bibinfo {year} {2023})}\BibitemShut {NoStop}%
\bibitem [{\citenamefont {Ren}\ \emph {et~al.}(2021)\citenamefont {Ren}, \citenamefont {Franke},\ and\ \citenamefont {Hughes}}]{ren2021quasinormal}%
  \BibitemOpen
  \bibfield  {author} {\bibinfo {author} {\bibfnamefont {J.}~\bibnamefont {Ren}}, \bibinfo {author} {\bibfnamefont {S.}~\bibnamefont {Franke}},\ and\ \bibinfo {author} {\bibfnamefont {S.}~\bibnamefont {Hughes}},\ }\bibfield  {title} {\bibinfo {title} {Quasinormal modes, local density of states, and classical purcell factors for coupled loss-gain resonators},\ }\href@noop {} {\bibfield  {journal} {\bibinfo  {journal} {Physical Review X}\ }\textbf {\bibinfo {volume} {11}},\ \bibinfo {pages} {041020} (\bibinfo {year} {2021})}\BibitemShut {NoStop}%
\bibitem [{\citenamefont {Lalanne}\ \emph {et~al.}(2019)\citenamefont {Lalanne}, \citenamefont {Yan}, \citenamefont {Gras}, \citenamefont {Sauvan}, \citenamefont {Hugonin}, \citenamefont {Besbes}, \citenamefont {Dem{\'e}sy}, \citenamefont {Truong}, \citenamefont {Gralak}, \citenamefont {Zolla} \emph {et~al.}}]{lalanne2019quasinormal}%
  \BibitemOpen
  \bibfield  {author} {\bibinfo {author} {\bibfnamefont {P.}~\bibnamefont {Lalanne}}, \bibinfo {author} {\bibfnamefont {W.}~\bibnamefont {Yan}}, \bibinfo {author} {\bibfnamefont {A.}~\bibnamefont {Gras}}, \bibinfo {author} {\bibfnamefont {C.}~\bibnamefont {Sauvan}}, \bibinfo {author} {\bibfnamefont {J.-P.}\ \bibnamefont {Hugonin}}, \bibinfo {author} {\bibfnamefont {M.}~\bibnamefont {Besbes}}, \bibinfo {author} {\bibfnamefont {G.}~\bibnamefont {Dem{\'e}sy}}, \bibinfo {author} {\bibfnamefont {M.}~\bibnamefont {Truong}}, \bibinfo {author} {\bibfnamefont {B.}~\bibnamefont {Gralak}}, \bibinfo {author} {\bibfnamefont {F.}~\bibnamefont {Zolla}}, \emph {et~al.},\ }\bibfield  {title} {\bibinfo {title} {Quasinormal mode solvers for resonators with dispersive materials},\ }\href@noop {} {\bibfield  {journal} {\bibinfo  {journal} {JOSA A}\ }\textbf {\bibinfo {volume} {36}},\ \bibinfo {pages} {686} (\bibinfo {year} {2019})}\BibitemShut {NoStop}%
\bibitem [{\citenamefont {Kristensen}\ \emph {et~al.}(2020)\citenamefont {Kristensen}, \citenamefont {Herrmann}, \citenamefont {Intravaia},\ and\ \citenamefont {Busch}}]{kristensen2020modeling}%
  \BibitemOpen
  \bibfield  {author} {\bibinfo {author} {\bibfnamefont {P.~T.}\ \bibnamefont {Kristensen}}, \bibinfo {author} {\bibfnamefont {K.}~\bibnamefont {Herrmann}}, \bibinfo {author} {\bibfnamefont {F.}~\bibnamefont {Intravaia}},\ and\ \bibinfo {author} {\bibfnamefont {K.}~\bibnamefont {Busch}},\ }\bibfield  {title} {\bibinfo {title} {Modeling electromagnetic resonators using quasinormal modes},\ }\href@noop {} {\bibfield  {journal} {\bibinfo  {journal} {Advances in Optics and Photonics}\ }\textbf {\bibinfo {volume} {12}},\ \bibinfo {pages} {612} (\bibinfo {year} {2020})}\BibitemShut {NoStop}%
\bibitem [{\citenamefont {Alam}\ \emph {et~al.}(2016)\citenamefont {Alam}, \citenamefont {De~Leon},\ and\ \citenamefont {Boyd}}]{alam2016large}%
  \BibitemOpen
  \bibfield  {author} {\bibinfo {author} {\bibfnamefont {M.~Z.}\ \bibnamefont {Alam}}, \bibinfo {author} {\bibfnamefont {I.}~\bibnamefont {De~Leon}},\ and\ \bibinfo {author} {\bibfnamefont {R.~W.}\ \bibnamefont {Boyd}},\ }\bibfield  {title} {\bibinfo {title} {Large optical nonlinearity of indium tin oxide in its epsilon-near-zero region},\ }\href@noop {} {\bibfield  {journal} {\bibinfo  {journal} {Science}\ }\textbf {\bibinfo {volume} {352}},\ \bibinfo {pages} {795} (\bibinfo {year} {2016})}\BibitemShut {NoStop}%
\bibitem [{\citenamefont {Capretti}\ \emph {et~al.}(2015{\natexlab{a}})\citenamefont {Capretti}, \citenamefont {Wang}, \citenamefont {Engheta},\ and\ \citenamefont {Dal~Negro}}]{capretti2015enhanced}%
  \BibitemOpen
  \bibfield  {author} {\bibinfo {author} {\bibfnamefont {A.}~\bibnamefont {Capretti}}, \bibinfo {author} {\bibfnamefont {Y.}~\bibnamefont {Wang}}, \bibinfo {author} {\bibfnamefont {N.}~\bibnamefont {Engheta}},\ and\ \bibinfo {author} {\bibfnamefont {L.}~\bibnamefont {Dal~Negro}},\ }\bibfield  {title} {\bibinfo {title} {Enhanced third-harmonic generation in si-compatible epsilon-near-zero indium tin oxide nanolayers},\ }\href@noop {} {\bibfield  {journal} {\bibinfo  {journal} {Optics letters}\ }\textbf {\bibinfo {volume} {40}},\ \bibinfo {pages} {1500} (\bibinfo {year} {2015}{\natexlab{a}})}\BibitemShut {NoStop}%
\bibitem [{\citenamefont {Reshef}\ \emph {et~al.}(2017)\citenamefont {Reshef}, \citenamefont {Giese}, \citenamefont {Alam}, \citenamefont {De~Leon}, \citenamefont {Upham},\ and\ \citenamefont {Boyd}}]{reshef2017beyond}%
  \BibitemOpen
  \bibfield  {author} {\bibinfo {author} {\bibfnamefont {O.}~\bibnamefont {Reshef}}, \bibinfo {author} {\bibfnamefont {E.}~\bibnamefont {Giese}}, \bibinfo {author} {\bibfnamefont {M.~Z.}\ \bibnamefont {Alam}}, \bibinfo {author} {\bibfnamefont {I.}~\bibnamefont {De~Leon}}, \bibinfo {author} {\bibfnamefont {J.}~\bibnamefont {Upham}},\ and\ \bibinfo {author} {\bibfnamefont {R.~W.}\ \bibnamefont {Boyd}},\ }\bibfield  {title} {\bibinfo {title} {Beyond the perturbative description of the nonlinear optical response of low-index materials},\ }\href@noop {} {\bibfield  {journal} {\bibinfo  {journal} {Optics letters}\ }\textbf {\bibinfo {volume} {42}},\ \bibinfo {pages} {3225} (\bibinfo {year} {2017})}\BibitemShut {NoStop}%
\bibitem [{\citenamefont {Reshef}\ \emph {et~al.}(2019)\citenamefont {Reshef}, \citenamefont {De~Leon}, \citenamefont {Alam},\ and\ \citenamefont {Boyd}}]{reshef2019nonlinear}%
  \BibitemOpen
  \bibfield  {author} {\bibinfo {author} {\bibfnamefont {O.}~\bibnamefont {Reshef}}, \bibinfo {author} {\bibfnamefont {I.}~\bibnamefont {De~Leon}}, \bibinfo {author} {\bibfnamefont {M.~Z.}\ \bibnamefont {Alam}},\ and\ \bibinfo {author} {\bibfnamefont {R.~W.}\ \bibnamefont {Boyd}},\ }\bibfield  {title} {\bibinfo {title} {Nonlinear optical effects in epsilon-near-zero media},\ }\href@noop {} {\bibfield  {journal} {\bibinfo  {journal} {Nature Reviews Materials}\ }\textbf {\bibinfo {volume} {4}},\ \bibinfo {pages} {535} (\bibinfo {year} {2019})}\BibitemShut {NoStop}%
\bibitem [{\citenamefont {Shubitidze}\ \emph {et~al.}(2024)\citenamefont {Shubitidze}, \citenamefont {Britton},\ and\ \citenamefont {Dal~Negro}}]{shubitidze2024enhanced}%
  \BibitemOpen
  \bibfield  {author} {\bibinfo {author} {\bibfnamefont {T.}~\bibnamefont {Shubitidze}}, \bibinfo {author} {\bibfnamefont {W.~A.}\ \bibnamefont {Britton}},\ and\ \bibinfo {author} {\bibfnamefont {L.}~\bibnamefont {Dal~Negro}},\ }\bibfield  {title} {\bibinfo {title} {Enhanced nonlinearity of epsilon-near-zero indium tin oxide nanolayers with tamm plasmon-polariton states},\ }\href@noop {} {\bibfield  {journal} {\bibinfo  {journal} {Advanced Optical Materials}\ }\textbf {\bibinfo {volume} {12}},\ \bibinfo {pages} {2301669} (\bibinfo {year} {2024})}\BibitemShut {NoStop}%
\bibitem [{\citenamefont {Tamashevich}\ \emph {et~al.}(2024)\citenamefont {Tamashevich}, \citenamefont {Shubitidze}, \citenamefont {Dal~Negro},\ and\ \citenamefont {Ornigotti}}]{tamashevich2024field}%
  \BibitemOpen
  \bibfield  {author} {\bibinfo {author} {\bibfnamefont {Y.}~\bibnamefont {Tamashevich}}, \bibinfo {author} {\bibfnamefont {T.}~\bibnamefont {Shubitidze}}, \bibinfo {author} {\bibfnamefont {L.}~\bibnamefont {Dal~Negro}},\ and\ \bibinfo {author} {\bibfnamefont {M.}~\bibnamefont {Ornigotti}},\ }\bibfield  {title} {\bibinfo {title} {Field theory description of the non-perturbative optical nonlinearity of epsilon-near-zero media},\ }\href@noop {} {\bibfield  {journal} {\bibinfo  {journal} {APL Photonics}\ }\textbf {\bibinfo {volume} {9}},\ \bibinfo {pages} {016105} (\bibinfo {year} {2024})}\BibitemShut {NoStop}%
\bibitem [{\citenamefont {Shubitidze}\ \emph {et~al.}(2025)\citenamefont {Shubitidze}, \citenamefont {Chawla},\ and\ \citenamefont {Dal~Negro}}]{shubitidze2025enhancement}%
  \BibitemOpen
  \bibfield  {author} {\bibinfo {author} {\bibfnamefont {T.}~\bibnamefont {Shubitidze}}, \bibinfo {author} {\bibfnamefont {S.}~\bibnamefont {Chawla}},\ and\ \bibinfo {author} {\bibfnamefont {L.}~\bibnamefont {Dal~Negro}},\ }\bibfield  {title} {\bibinfo {title} {Enhancement of the third harmonic generation efficiency of ito nanolayers coupled to tamm plasmon polaritons},\ }\href@noop {} {\bibfield  {journal} {\bibinfo  {journal} {APL Photonics}\ }\textbf {\bibinfo {volume} {10}} (\bibinfo {year} {2025})}\BibitemShut {NoStop}%
\bibitem [{\citenamefont {Capretti}\ \emph {et~al.}(2015{\natexlab{b}})\citenamefont {Capretti}, \citenamefont {Wang}, \citenamefont {Engheta},\ and\ \citenamefont {Dal~Negro}}]{capretti2015comparative}%
  \BibitemOpen
  \bibfield  {author} {\bibinfo {author} {\bibfnamefont {A.}~\bibnamefont {Capretti}}, \bibinfo {author} {\bibfnamefont {Y.}~\bibnamefont {Wang}}, \bibinfo {author} {\bibfnamefont {N.}~\bibnamefont {Engheta}},\ and\ \bibinfo {author} {\bibfnamefont {L.}~\bibnamefont {Dal~Negro}},\ }\bibfield  {title} {\bibinfo {title} {Comparative study of second-harmonic generation from epsilon-near-zero indium tin oxide and titanium nitride nanolayers excited in the near-infrared spectral range},\ }\href@noop {} {\bibfield  {journal} {\bibinfo  {journal} {Acs Photonics}\ }\textbf {\bibinfo {volume} {2}},\ \bibinfo {pages} {1584} (\bibinfo {year} {2015}{\natexlab{b}})}\BibitemShut {NoStop}%
\bibitem [{\citenamefont {Chang}\ \emph {et~al.}(2014)\citenamefont {Chang}, \citenamefont {Vuleti{\'c}},\ and\ \citenamefont {Lukin}}]{chang2014quantum}%
  \BibitemOpen
  \bibfield  {author} {\bibinfo {author} {\bibfnamefont {D.~E.}\ \bibnamefont {Chang}}, \bibinfo {author} {\bibfnamefont {V.}~\bibnamefont {Vuleti{\'c}}},\ and\ \bibinfo {author} {\bibfnamefont {M.~D.}\ \bibnamefont {Lukin}},\ }\bibfield  {title} {\bibinfo {title} {Quantum nonlinear optics—photon by photon},\ }\href@noop {} {\bibfield  {journal} {\bibinfo  {journal} {Nature Photonics}\ }\textbf {\bibinfo {volume} {8}},\ \bibinfo {pages} {685} (\bibinfo {year} {2014})}\BibitemShut {NoStop}%
\bibitem [{\citenamefont {Jensen}\ and\ \citenamefont {Sigmund}(2011)}]{jensen2011topology}%
  \BibitemOpen
  \bibfield  {author} {\bibinfo {author} {\bibfnamefont {J.~S.}\ \bibnamefont {Jensen}}\ and\ \bibinfo {author} {\bibfnamefont {O.}~\bibnamefont {Sigmund}},\ }\bibfield  {title} {\bibinfo {title} {Topology optimization for nano-photonics},\ }\href@noop {} {\bibfield  {journal} {\bibinfo  {journal} {Laser \& Photonics Reviews}\ }\textbf {\bibinfo {volume} {5}},\ \bibinfo {pages} {308} (\bibinfo {year} {2011})}\BibitemShut {NoStop}%
\bibitem [{\citenamefont {Choi}\ \emph {et~al.}(2017)\citenamefont {Choi}, \citenamefont {Heuck},\ and\ \citenamefont {Englund}}]{choi2017self}%
  \BibitemOpen
  \bibfield  {author} {\bibinfo {author} {\bibfnamefont {H.}~\bibnamefont {Choi}}, \bibinfo {author} {\bibfnamefont {M.}~\bibnamefont {Heuck}},\ and\ \bibinfo {author} {\bibfnamefont {D.}~\bibnamefont {Englund}},\ }\bibfield  {title} {\bibinfo {title} {Self-similar nanocavity design with ultrasmall mode volume for single-photon nonlinearities},\ }\href@noop {} {\bibfield  {journal} {\bibinfo  {journal} {Physical review letters}\ }\textbf {\bibinfo {volume} {118}},\ \bibinfo {pages} {223605} (\bibinfo {year} {2017})}\BibitemShut {NoStop}%
\bibitem [{\citenamefont {Albrechtsen}\ \emph {et~al.}(2022)\citenamefont {Albrechtsen}, \citenamefont {Vosoughi~Lahijani}, \citenamefont {Christiansen}, \citenamefont {Nguyen}, \citenamefont {Casses}, \citenamefont {Hansen}, \citenamefont {Stenger}, \citenamefont {Sigmund}, \citenamefont {Jansen}, \citenamefont {M{\o}rk} \emph {et~al.}}]{albrechtsen2022nanometer}%
  \BibitemOpen
  \bibfield  {author} {\bibinfo {author} {\bibfnamefont {M.}~\bibnamefont {Albrechtsen}}, \bibinfo {author} {\bibfnamefont {B.}~\bibnamefont {Vosoughi~Lahijani}}, \bibinfo {author} {\bibfnamefont {R.~E.}\ \bibnamefont {Christiansen}}, \bibinfo {author} {\bibfnamefont {V.~T.~H.}\ \bibnamefont {Nguyen}}, \bibinfo {author} {\bibfnamefont {L.~N.}\ \bibnamefont {Casses}}, \bibinfo {author} {\bibfnamefont {S.~E.}\ \bibnamefont {Hansen}}, \bibinfo {author} {\bibfnamefont {N.}~\bibnamefont {Stenger}}, \bibinfo {author} {\bibfnamefont {O.}~\bibnamefont {Sigmund}}, \bibinfo {author} {\bibfnamefont {H.}~\bibnamefont {Jansen}}, \bibinfo {author} {\bibfnamefont {J.}~\bibnamefont {M{\o}rk}}, \emph {et~al.},\ }\bibfield  {title} {\bibinfo {title} {Nanometer-scale photon confinement in topology-optimized dielectric cavities},\ }\href@noop {} {\bibfield  {journal} {\bibinfo  {journal} {Nature Communications}\ }\textbf {\bibinfo {volume} {13}},\ \bibinfo {pages} {6281} (\bibinfo {year} {2022})}\BibitemShut {NoStop}%
\bibitem [{\citenamefont {Christiansen}\ and\ \citenamefont {Sigmund}(2021)}]{christiansen2021inverse}%
  \BibitemOpen
  \bibfield  {author} {\bibinfo {author} {\bibfnamefont {R.~E.}\ \bibnamefont {Christiansen}}\ and\ \bibinfo {author} {\bibfnamefont {O.}~\bibnamefont {Sigmund}},\ }\bibfield  {title} {\bibinfo {title} {Inverse design in photonics by topology optimization: tutorial},\ }\href@noop {} {\bibfield  {journal} {\bibinfo  {journal} {JOSA B}\ }\textbf {\bibinfo {volume} {38}},\ \bibinfo {pages} {496} (\bibinfo {year} {2021})}\BibitemShut {NoStop}%
\bibitem [{\citenamefont {Mignuzzi}\ \emph {et~al.}(2019)\citenamefont {Mignuzzi}, \citenamefont {Vezzoli}, \citenamefont {Horsley}, \citenamefont {Barnes}, \citenamefont {Maier},\ and\ \citenamefont {Sapienza}}]{mignuzzi2019nanoscale}%
  \BibitemOpen
  \bibfield  {author} {\bibinfo {author} {\bibfnamefont {S.}~\bibnamefont {Mignuzzi}}, \bibinfo {author} {\bibfnamefont {S.}~\bibnamefont {Vezzoli}}, \bibinfo {author} {\bibfnamefont {S.~A.}\ \bibnamefont {Horsley}}, \bibinfo {author} {\bibfnamefont {W.~L.}\ \bibnamefont {Barnes}}, \bibinfo {author} {\bibfnamefont {S.~A.}\ \bibnamefont {Maier}},\ and\ \bibinfo {author} {\bibfnamefont {R.}~\bibnamefont {Sapienza}},\ }\bibfield  {title} {\bibinfo {title} {Nanoscale design of the local density of optical states},\ }\href@noop {} {\bibfield  {journal} {\bibinfo  {journal} {Nano Letters}\ }\textbf {\bibinfo {volume} {19}},\ \bibinfo {pages} {1613} (\bibinfo {year} {2019})}\BibitemShut {NoStop}%
\bibitem [{\citenamefont {Molesky}\ \emph {et~al.}(2018)\citenamefont {Molesky}, \citenamefont {Lin}, \citenamefont {Piggott}, \citenamefont {Jin}, \citenamefont {Vuckovi{\'c}},\ and\ \citenamefont {Rodriguez}}]{molesky2018inverse}%
  \BibitemOpen
  \bibfield  {author} {\bibinfo {author} {\bibfnamefont {S.}~\bibnamefont {Molesky}}, \bibinfo {author} {\bibfnamefont {Z.}~\bibnamefont {Lin}}, \bibinfo {author} {\bibfnamefont {A.~Y.}\ \bibnamefont {Piggott}}, \bibinfo {author} {\bibfnamefont {W.}~\bibnamefont {Jin}}, \bibinfo {author} {\bibfnamefont {J.}~\bibnamefont {Vuckovi{\'c}}},\ and\ \bibinfo {author} {\bibfnamefont {A.~W.}\ \bibnamefont {Rodriguez}},\ }\bibfield  {title} {\bibinfo {title} {Inverse design in nanophotonics},\ }\href@noop {} {\bibfield  {journal} {\bibinfo  {journal} {Nature Photonics}\ }\textbf {\bibinfo {volume} {12}},\ \bibinfo {pages} {659} (\bibinfo {year} {2018})}\BibitemShut {NoStop}%
\bibitem [{\citenamefont {Sehmi}\ \emph {et~al.}(2020)\citenamefont {Sehmi}, \citenamefont {Langbein},\ and\ \citenamefont {Muljarov}}]{resStates2}%
  \BibitemOpen
  \bibfield  {author} {\bibinfo {author} {\bibfnamefont {H.~S.}\ \bibnamefont {Sehmi}}, \bibinfo {author} {\bibfnamefont {W.}~\bibnamefont {Langbein}},\ and\ \bibinfo {author} {\bibfnamefont {E.~A.}\ \bibnamefont {Muljarov}},\ }\bibfield  {title} {\bibinfo {title} {Applying the resonant state expansion to realistic materials with frequency dispersion},\ }\href@noop {} {\bibfield  {journal} {\bibinfo  {journal} {Phys. Rev. B}\ }\textbf {\bibinfo {volume} {101}},\ \bibinfo {pages} {045304} (\bibinfo {year} {2020})}\BibitemShut {NoStop}%
\bibitem [{\citenamefont {Wu}\ \emph {et~al.}(2021{\natexlab{a}})\citenamefont {Wu}, \citenamefont {Gurioli},\ and\ \citenamefont {Lalanne}}]{wu2021nanoscale}%
  \BibitemOpen
  \bibfield  {author} {\bibinfo {author} {\bibfnamefont {T.}~\bibnamefont {Wu}}, \bibinfo {author} {\bibfnamefont {M.}~\bibnamefont {Gurioli}},\ and\ \bibinfo {author} {\bibfnamefont {P.}~\bibnamefont {Lalanne}},\ }\bibfield  {title} {\bibinfo {title} {Nanoscale light confinement: the q’s and v’s},\ }\href@noop {} {\bibfield  {journal} {\bibinfo  {journal} {ACS photonics}\ }\textbf {\bibinfo {volume} {8}},\ \bibinfo {pages} {1522} (\bibinfo {year} {2021}{\natexlab{a}})}\BibitemShut {NoStop}%
\bibitem [{\citenamefont {Britton}\ \emph {et~al.}(2022)\citenamefont {Britton}, \citenamefont {Sgrignuoli},\ and\ \citenamefont {Dal~Negro}}]{britton2022structure}%
  \BibitemOpen
  \bibfield  {author} {\bibinfo {author} {\bibfnamefont {W.~A.}\ \bibnamefont {Britton}}, \bibinfo {author} {\bibfnamefont {F.}~\bibnamefont {Sgrignuoli}},\ and\ \bibinfo {author} {\bibfnamefont {L.}~\bibnamefont {Dal~Negro}},\ }\bibfield  {title} {\bibinfo {title} {Structure-dependent optical nonlinearity of indium tin oxide},\ }\href@noop {} {\bibfield  {journal} {\bibinfo  {journal} {Applied Physics Letters}\ }\textbf {\bibinfo {volume} {120}} (\bibinfo {year} {2022})}\BibitemShut {NoStop}%
\bibitem [{\citenamefont {Dung}\ \emph {et~al.}(1998)\citenamefont {Dung}, \citenamefont {Kn{\"o}ll},\ and\ \citenamefont {Welsch}}]{dung1998three}%
  \BibitemOpen
  \bibfield  {author} {\bibinfo {author} {\bibfnamefont {H.~T.}\ \bibnamefont {Dung}}, \bibinfo {author} {\bibfnamefont {L.}~\bibnamefont {Kn{\"o}ll}},\ and\ \bibinfo {author} {\bibfnamefont {D.-G.}\ \bibnamefont {Welsch}},\ }\bibfield  {title} {\bibinfo {title} {Three-dimensional quantization of the electromagnetic field in dispersive and absorbing inhomogeneous dielectrics},\ }\href@noop {} {\bibfield  {journal} {\bibinfo  {journal} {Physical Review A}\ }\textbf {\bibinfo {volume} {57}},\ \bibinfo {pages} {3931} (\bibinfo {year} {1998})}\BibitemShut {NoStop}%
\bibitem [{\citenamefont {Drezet}(2017{\natexlab{a}})}]{drezet2017equivalence}%
  \BibitemOpen
  \bibfield  {author} {\bibinfo {author} {\bibfnamefont {A.}~\bibnamefont {Drezet}},\ }\bibfield  {title} {\bibinfo {title} {Equivalence between the hamiltonian and langevin noise descriptions of plasmon polaritons in a dispersive and lossy inhomogeneous medium},\ }\href@noop {} {\bibfield  {journal} {\bibinfo  {journal} {Physical Review A}\ }\textbf {\bibinfo {volume} {96}},\ \bibinfo {pages} {033849} (\bibinfo {year} {2017}{\natexlab{a}})}\BibitemShut {NoStop}%
\bibitem [{\citenamefont {Drezet}(2017{\natexlab{b}})}]{drezet2017quantizing}%
  \BibitemOpen
  \bibfield  {author} {\bibinfo {author} {\bibfnamefont {A.}~\bibnamefont {Drezet}},\ }\bibfield  {title} {\bibinfo {title} {Quantizing polaritons in inhomogeneous dissipative systems},\ }\href@noop {} {\bibfield  {journal} {\bibinfo  {journal} {Physical Review A}\ }\textbf {\bibinfo {volume} {95}},\ \bibinfo {pages} {023831} (\bibinfo {year} {2017}{\natexlab{b}})}\BibitemShut {NoStop}%
\bibitem [{\citenamefont {Boyd}(2020)}]{boyd_nonlinear_2020}%
  \BibitemOpen
  \bibfield  {author} {\bibinfo {author} {\bibfnamefont {R.~W.}\ \bibnamefont {Boyd}},\ }\href {https://doi.org/10.1016/C2015-0-05510-1} {{\selectlanguage {english}\emph {\bibinfo {title} {Nonlinear optics}}}},\ \bibinfo {edition} {fourth edition}\ ed.\ (\bibinfo  {publisher} {Elsevier, AP Academic Press},\ \bibinfo {address} {London},\ \bibinfo {year} {2020})\BibitemShut {NoStop}%
\bibitem [{\citenamefont {Schubert}\ and\ \citenamefont {Wilhelmi}(1986)}]{wilhelmi}%
  \BibitemOpen
  \bibfield  {author} {\bibinfo {author} {\bibfnamefont {M.}~\bibnamefont {Schubert}}\ and\ \bibinfo {author} {\bibfnamefont {B.}~\bibnamefont {Wilhelmi}},\ }\href@noop {} {\emph {\bibinfo {title} {Nonlinear Optics and Quantum Electronics}}}\ (\bibinfo  {publisher} {Wiley-Interscience},\ \bibinfo {year} {1986})\BibitemShut {NoStop}%
\bibitem [{\citenamefont {Srednicki}(2007)}]{srednicki}%
  \BibitemOpen
  \bibfield  {author} {\bibinfo {author} {\bibfnamefont {M.}~\bibnamefont {Srednicki}},\ }\href@noop {} {\emph {\bibinfo {title} {Quantum Field Theory}}}\ (\bibinfo  {publisher} {Cambridge University Press},\ \bibinfo {year} {2007})\BibitemShut {NoStop}%
\bibitem [{\citenamefont {Haroche}\ and\ \citenamefont {Raimond}(2006)}]{haroche}%
  \BibitemOpen
  \bibfield  {author} {\bibinfo {author} {\bibfnamefont {S.}~\bibnamefont {Haroche}}\ and\ \bibinfo {author} {\bibfnamefont {J.~M.}\ \bibnamefont {Raimond}},\ }\href@noop {} {\emph {\bibinfo {title} {Exploring the quantum. Atom, cavities, and photons}}}\ (\bibinfo  {publisher} {Oxford University Press},\ \bibinfo {year} {2006})\BibitemShut {NoStop}%
\bibitem [{\citenamefont {Kirchmair}\ \emph {et~al.}(2013)\citenamefont {Kirchmair}, \citenamefont {Vlastakis}, \citenamefont {Leghtas}, \citenamefont {Nigg}, \citenamefont {Paik}, \citenamefont {Ginossar}, \citenamefont {Mirrahimi}, \citenamefont {Frunzion}, \citenamefont {Girvin},\ and\ \citenamefont {Schoelkopf}}]{kirchmair}%
  \BibitemOpen
  \bibfield  {author} {\bibinfo {author} {\bibfnamefont {G.}~\bibnamefont {Kirchmair}}, \bibinfo {author} {\bibfnamefont {B.}~\bibnamefont {Vlastakis}}, \bibinfo {author} {\bibfnamefont {Z.}~\bibnamefont {Leghtas}}, \bibinfo {author} {\bibfnamefont {S.~E.}\ \bibnamefont {Nigg}}, \bibinfo {author} {\bibfnamefont {H.}~\bibnamefont {Paik}}, \bibinfo {author} {\bibfnamefont {E.}~\bibnamefont {Ginossar}}, \bibinfo {author} {\bibfnamefont {M.}~\bibnamefont {Mirrahimi}}, \bibinfo {author} {\bibfnamefont {L.}~\bibnamefont {Frunzion}}, \bibinfo {author} {\bibfnamefont {S.~M.}\ \bibnamefont {Girvin}},\ and\ \bibinfo {author} {\bibfnamefont {R.~J.}\ \bibnamefont {Schoelkopf}},\ }\bibfield  {title} {\bibinfo {title} {Observation of quantum state collapse and revival due to the single-photon kerr effect},\ }\href@noop {} {\bibfield  {journal} {\bibinfo  {journal} {Nature}\ }\textbf {\bibinfo {volume} {495}},\ \bibinfo {pages} {205} (\bibinfo {year} {2013})}\BibitemShut {NoStop}%
\bibitem [{\citenamefont {Drummond}\ and\ \citenamefont {Walls}(1980)}]{drummondKerr}%
  \BibitemOpen
  \bibfield  {author} {\bibinfo {author} {\bibfnamefont {P.~D.}\ \bibnamefont {Drummond}}\ and\ \bibinfo {author} {\bibfnamefont {D.~F.}\ \bibnamefont {Walls}},\ }\bibfield  {title} {\bibinfo {title} {Quantum theory of optical bistability i: nonlinear polarisability model},\ }\href@noop {} {\bibfield  {journal} {\bibinfo  {journal} {J. Phys A: Mat, Gen.}\ }\textbf {\bibinfo {volume} {13}},\ \bibinfo {pages} {725} (\bibinfo {year} {1980})}\BibitemShut {NoStop}%
\bibitem [{\citenamefont {Bialynicki-Birula}\ and\ \citenamefont {Zofia}(2017)}]{birulaPhotons}%
  \BibitemOpen
  \bibfield  {author} {\bibinfo {author} {\bibfnamefont {I.}~\bibnamefont {Bialynicki-Birula}}\ and\ \bibinfo {author} {\bibfnamefont {B.-B.}\ \bibnamefont {Zofia}},\ }\bibfield  {title} {\bibinfo {title} {Quantum-mechanical description of optical beams},\ }\href@noop {} {\bibfield  {journal} {\bibinfo  {journal} {J. Opt.}\ }\textbf {\bibinfo {volume} {19}},\ \bibinfo {pages} {125201} (\bibinfo {year} {2017})}\BibitemShut {NoStop}%
\bibitem [{\citenamefont {Breuer}\ and\ \citenamefont {Petruccione}(2007)}]{breuerQS}%
  \BibitemOpen
  \bibfield  {author} {\bibinfo {author} {\bibfnamefont {H.~P.}\ \bibnamefont {Breuer}}\ and\ \bibinfo {author} {\bibfnamefont {F.}~\bibnamefont {Petruccione}},\ }\href@noop {} {\emph {\bibinfo {title} {The theory of open quantum systems}}}\ (\bibinfo  {publisher} {Oxford University Press},\ \bibinfo {year} {2007})\BibitemShut {NoStop}%
\bibitem [{\citenamefont {Doost}\ \emph {et~al.}(2014)\citenamefont {Doost}, \citenamefont {Langbein},\ and\ \citenamefont {Muljarov}}]{resStates}%
  \BibitemOpen
  \bibfield  {author} {\bibinfo {author} {\bibfnamefont {M.~B.}\ \bibnamefont {Doost}}, \bibinfo {author} {\bibfnamefont {W.}~\bibnamefont {Langbein}},\ and\ \bibinfo {author} {\bibfnamefont {E.~A.}\ \bibnamefont {Muljarov}},\ }\bibfield  {title} {\bibinfo {title} {Resonant-state expansion applied to three-dimensional open optical systems},\ }\href {https://doi.org/10.1103/PhysRevA.90.013834} {\bibfield  {journal} {\bibinfo  {journal} {Phys. Rev. A}\ }\textbf {\bibinfo {volume} {90}},\ \bibinfo {pages} {013834} (\bibinfo {year} {2014})}\BibitemShut {NoStop}%
\bibitem [{\citenamefont {Lobanov}\ \emph {et~al.}(2018)\citenamefont {Lobanov}, \citenamefont {Langbein},\ and\ \citenamefont {Muljarov}}]{lobanov2018resonant}%
  \BibitemOpen
  \bibfield  {author} {\bibinfo {author} {\bibfnamefont {S.~V.}\ \bibnamefont {Lobanov}}, \bibinfo {author} {\bibfnamefont {W.}~\bibnamefont {Langbein}},\ and\ \bibinfo {author} {\bibfnamefont {E.~A.}\ \bibnamefont {Muljarov}},\ }\bibfield  {title} {\bibinfo {title} {Resonant-state expansion of three-dimensional open optical systems: Light scattering},\ }\href@noop {} {\bibfield  {journal} {\bibinfo  {journal} {Physical Review A}\ }\textbf {\bibinfo {volume} {98}},\ \bibinfo {pages} {033820} (\bibinfo {year} {2018})}\BibitemShut {NoStop}%
\bibitem [{\citenamefont {Sauvan}\ \emph {et~al.}(2022{\natexlab{a}})\citenamefont {Sauvan}, \citenamefont {Wu}, \citenamefont {Zarouf}, \citenamefont {Muljarov},\ and\ \citenamefont {Lalanne}}]{sauvan2022normalization}%
  \BibitemOpen
  \bibfield  {author} {\bibinfo {author} {\bibfnamefont {C.}~\bibnamefont {Sauvan}}, \bibinfo {author} {\bibfnamefont {T.}~\bibnamefont {Wu}}, \bibinfo {author} {\bibfnamefont {R.}~\bibnamefont {Zarouf}}, \bibinfo {author} {\bibfnamefont {E.~A.}\ \bibnamefont {Muljarov}},\ and\ \bibinfo {author} {\bibfnamefont {P.}~\bibnamefont {Lalanne}},\ }\bibfield  {title} {\bibinfo {title} {Normalization, orthogonality, and completeness of quasinormal modes of open systems: the case of electromagnetism},\ }\href@noop {} {\bibfield  {journal} {\bibinfo  {journal} {Optics Express}\ }\textbf {\bibinfo {volume} {30}},\ \bibinfo {pages} {6846} (\bibinfo {year} {2022}{\natexlab{a}})}\BibitemShut {NoStop}%
\bibitem [{\citenamefont {Lalanne}\ \emph {et~al.}(2018)\citenamefont {Lalanne}, \citenamefont {Yan}, \citenamefont {Vynck}, \citenamefont {Sauvan},\ and\ \citenamefont {Hugonin}}]{lalanne2018light}%
  \BibitemOpen
  \bibfield  {author} {\bibinfo {author} {\bibfnamefont {P.}~\bibnamefont {Lalanne}}, \bibinfo {author} {\bibfnamefont {W.}~\bibnamefont {Yan}}, \bibinfo {author} {\bibfnamefont {K.}~\bibnamefont {Vynck}}, \bibinfo {author} {\bibfnamefont {C.}~\bibnamefont {Sauvan}},\ and\ \bibinfo {author} {\bibfnamefont {J.-P.}\ \bibnamefont {Hugonin}},\ }\bibfield  {title} {\bibinfo {title} {Light interaction with photonic and plasmonic resonances},\ }\href@noop {} {\bibfield  {journal} {\bibinfo  {journal} {Laser \& Photonics Reviews}\ }\textbf {\bibinfo {volume} {12}},\ \bibinfo {pages} {1700113} (\bibinfo {year} {2018})}\BibitemShut {NoStop}%
\bibitem [{\citenamefont {Jackson}(2021)}]{jackson2021classical}%
  \BibitemOpen
  \bibfield  {author} {\bibinfo {author} {\bibfnamefont {J.~D.}\ \bibnamefont {Jackson}},\ }\href@noop {} {\emph {\bibinfo {title} {Classical electrodynamics}}}\ (\bibinfo  {publisher} {John Wiley \& Sons},\ \bibinfo {year} {2021})\BibitemShut {NoStop}%
\bibitem [{\citenamefont {Olver}\ \emph {et~al.}(2010)\citenamefont {Olver}, \citenamefont {Loier}, \citenamefont {Boisvert},\ and\ \citenamefont {Clark}}]{nist}%
  \BibitemOpen
  \bibfield  {author} {\bibinfo {author} {\bibfnamefont {F.~W.~J.}\ \bibnamefont {Olver}}, \bibinfo {author} {\bibfnamefont {D.~W.}\ \bibnamefont {Loier}}, \bibinfo {author} {\bibfnamefont {R.~F.}\ \bibnamefont {Boisvert}},\ and\ \bibinfo {author} {\bibfnamefont {C.~W.}\ \bibnamefont {Clark}},\ }\href@noop {} {\emph {\bibinfo {title} {NIST Handbook of Mathematical Functions}}}\ (\bibinfo  {publisher} {Cambridge University Press},\ \bibinfo {year} {2010})\BibitemShut {NoStop}%
\bibitem [{\citenamefont {Bohren}\ and\ \citenamefont {Huffman}(1983)}]{bohren}%
  \BibitemOpen
  \bibfield  {author} {\bibinfo {author} {\bibfnamefont {C.~F.}\ \bibnamefont {Bohren}}\ and\ \bibinfo {author} {\bibfnamefont {D.~R.}\ \bibnamefont {Huffman}},\ }\href@noop {} {\emph {\bibinfo {title} {Absorption and scattering of light by small particles}}}\ (\bibinfo  {publisher} {Wiley},\ \bibinfo {year} {1983})\BibitemShut {NoStop}%
\bibitem [{\citenamefont {Wu}\ \emph {et~al.}(2021{\natexlab{b}})\citenamefont {Wu}, \citenamefont {Gurioli},\ and\ \citenamefont {Lalanne}}]{wu_nanoscale_2021}%
  \BibitemOpen
  \bibfield  {author} {\bibinfo {author} {\bibfnamefont {T.}~\bibnamefont {Wu}}, \bibinfo {author} {\bibfnamefont {M.}~\bibnamefont {Gurioli}},\ and\ \bibinfo {author} {\bibfnamefont {P.}~\bibnamefont {Lalanne}},\ }\bibfield  {title} {\bibinfo {title} {Nanoscale {Light} {Confinement}: the {Q}{\textquoteright}s and {V}{\textquoteright}s},\ }\href {https://doi.org/10.1021/acsphotonics.1c00336} {\bibfield  {journal} {\bibinfo  {journal} {ACS Photonics}\ }\textbf {\bibinfo {volume} {8}},\ \bibinfo {pages} {1522} (\bibinfo {year} {2021}{\natexlab{b}})},\ \bibinfo {note} {publisher: American Chemical Society}\BibitemShut {NoStop}%
\bibitem [{\citenamefont {Wu}\ \emph {et~al.}(2023)\citenamefont {Wu}, \citenamefont {Arrivault}, \citenamefont {Yan},\ and\ \citenamefont {Lalanne}}]{wu_modal_2023}%
  \BibitemOpen
  \bibfield  {author} {\bibinfo {author} {\bibfnamefont {T.}~\bibnamefont {Wu}}, \bibinfo {author} {\bibfnamefont {D.}~\bibnamefont {Arrivault}}, \bibinfo {author} {\bibfnamefont {W.}~\bibnamefont {Yan}},\ and\ \bibinfo {author} {\bibfnamefont {P.}~\bibnamefont {Lalanne}},\ }\bibfield  {title} {{\selectlanguage {english}\bibinfo {title} {Modal analysis of electromagnetic resonators: {User} guide for the {MAN} program}},\ }\href {https://doi.org/10.1016/j.cpc.2022.108627} {\bibfield  {journal} {\bibinfo  {journal} {Computer Physics Communications}\ }\textbf {\bibinfo {volume} {284}},\ \bibinfo {pages} {108627} (\bibinfo {year} {2023})}\BibitemShut {NoStop}%
\bibitem [{\citenamefont {Sauvan}\ \emph {et~al.}(2022{\natexlab{b}})\citenamefont {Sauvan}, \citenamefont {Wu}, \citenamefont {Zarouf}, \citenamefont {Muljarov},\ and\ \citenamefont {Lalanne}}]{sauvan_normalization_2022}%
  \BibitemOpen
  \bibfield  {author} {\bibinfo {author} {\bibfnamefont {C.}~\bibnamefont {Sauvan}}, \bibinfo {author} {\bibfnamefont {T.}~\bibnamefont {Wu}}, \bibinfo {author} {\bibfnamefont {R.}~\bibnamefont {Zarouf}}, \bibinfo {author} {\bibfnamefont {E.~A.}\ \bibnamefont {Muljarov}},\ and\ \bibinfo {author} {\bibfnamefont {P.}~\bibnamefont {Lalanne}},\ }\bibfield  {title} {{\selectlanguage {english}\bibinfo {title} {Normalization, orthogonality, and completeness of quasinormal modes of open systems: the case of electromagnetism [{Invited}]}},\ }\href {https://doi.org/10.1364/OE.443656} {\bibfield  {journal} {\bibinfo  {journal} {Optics Express}\ }\textbf {\bibinfo {volume} {30}},\ \bibinfo {pages} {6846} (\bibinfo {year} {2022}{\natexlab{b}})},\ \bibinfo {note} {publisher: Optica Publishing Group}\BibitemShut {NoStop}%
\bibitem [{\citenamefont {Wu}\ and\ \citenamefont {Lalanne}(2021)}]{wu_qnmnonreciprocal_resonators_2021}%
  \BibitemOpen
  \bibfield  {author} {\bibinfo {author} {\bibfnamefont {T.}~\bibnamefont {Wu}}\ and\ \bibinfo {author} {\bibfnamefont {P.}~\bibnamefont {Lalanne}},\ }\href {https://doi.org/10.48550/arXiv.2106.05502} {\bibinfo {title} {{QNMnonreciprocal}\_resonators: an openly available toolbox for computing the {QuasiNormal} {Modes} of nonreciprocal resonators}} (\bibinfo {year} {2021}),\ \bibinfo {note} {arXiv:2106.05502 [physics]}\BibitemShut {NoStop}%
\bibitem [{\citenamefont {{COMSOL AB}}(2024)}]{comsol_sftw}%
  \BibitemOpen
  \bibfield  {author} {\bibinfo {author} {\bibnamefont {{COMSOL AB}}},\ }\href {https://www.comsol.com} {\bibinfo {title} {{COMSOL} {Multiphysics®} {v. 6.0,} {Stockholm,} {Sweden}}} (\bibinfo {year} {2024})\BibitemShut {NoStop}%
\bibitem [{\citenamefont {Cleri}\ \emph {et~al.}(2021)\citenamefont {Cleri}, \citenamefont {Tomko}, \citenamefont {Quiambao-Tomko}, \citenamefont {Imperatore}, \citenamefont {Zhu}, \citenamefont {Nolen}, \citenamefont {Nordlander}, \citenamefont {Caldwell}, \citenamefont {Mao}, \citenamefont {Giebink} \emph {et~al.}}]{cleri2021mid}%
  \BibitemOpen
  \bibfield  {author} {\bibinfo {author} {\bibfnamefont {A.}~\bibnamefont {Cleri}}, \bibinfo {author} {\bibfnamefont {J.}~\bibnamefont {Tomko}}, \bibinfo {author} {\bibfnamefont {K.}~\bibnamefont {Quiambao-Tomko}}, \bibinfo {author} {\bibfnamefont {M.~V.}\ \bibnamefont {Imperatore}}, \bibinfo {author} {\bibfnamefont {Y.}~\bibnamefont {Zhu}}, \bibinfo {author} {\bibfnamefont {J.~R.}\ \bibnamefont {Nolen}}, \bibinfo {author} {\bibfnamefont {J.}~\bibnamefont {Nordlander}}, \bibinfo {author} {\bibfnamefont {J.~D.}\ \bibnamefont {Caldwell}}, \bibinfo {author} {\bibfnamefont {Z.}~\bibnamefont {Mao}}, \bibinfo {author} {\bibfnamefont {N.~C.}\ \bibnamefont {Giebink}}, \emph {et~al.},\ }\bibfield  {title} {\bibinfo {title} {Mid-wave to near-ir optoelectronic properties and epsilon-near-zero behavior in indium-doped cadmium oxide},\ }\href@noop {} {\bibfield  {journal} {\bibinfo  {journal} {Physical Review Materials}\ }\textbf {\bibinfo {volume} {5}},\ \bibinfo {pages} {035202} (\bibinfo {year} {2021})}\BibitemShut
  {NoStop}%
\bibitem [{\citenamefont {Schrecengost}\ \emph {et~al.}(2024)\citenamefont {Schrecengost}, \citenamefont {Cleri}, \citenamefont {Tolchin}, \citenamefont {Mohan}, \citenamefont {Murphy}, \citenamefont {Adamkovic}, \citenamefont {Grede}, \citenamefont {Imperatore}, \citenamefont {Hopkins}, \citenamefont {Maria} \emph {et~al.}}]{schrecengost2024large}%
  \BibitemOpen
  \bibfield  {author} {\bibinfo {author} {\bibfnamefont {J.~R.}\ \bibnamefont {Schrecengost}}, \bibinfo {author} {\bibfnamefont {A.~J.}\ \bibnamefont {Cleri}}, \bibinfo {author} {\bibfnamefont {M.~J.}\ \bibnamefont {Tolchin}}, \bibinfo {author} {\bibfnamefont {R.}~\bibnamefont {Mohan}}, \bibinfo {author} {\bibfnamefont {J.~P.}\ \bibnamefont {Murphy}}, \bibinfo {author} {\bibfnamefont {S.}~\bibnamefont {Adamkovic}}, \bibinfo {author} {\bibfnamefont {A.~J.}\ \bibnamefont {Grede}}, \bibinfo {author} {\bibfnamefont {M.~V.}\ \bibnamefont {Imperatore}}, \bibinfo {author} {\bibfnamefont {P.~E.}\ \bibnamefont {Hopkins}}, \bibinfo {author} {\bibfnamefont {J.-P.}\ \bibnamefont {Maria}}, \emph {et~al.},\ }\bibfield  {title} {\bibinfo {title} {Large mid-infrared magneto-optic response from doped cadmium oxide at its epsilon-near-zero frequency},\ }\href@noop {} {\bibfield  {journal} {\bibinfo  {journal} {Advanced Optical Materials}\ }\textbf {\bibinfo {volume} {12}},\ \bibinfo {pages} {2400803} (\bibinfo {year}
  {2024})}\BibitemShut {NoStop}%
\bibitem [{\citenamefont {Stratton}(2010)}]{stratton}%
  \BibitemOpen
  \bibfield  {author} {\bibinfo {author} {\bibfnamefont {J.~A.}\ \bibnamefont {Stratton}},\ }\href@noop {} {\emph {\bibinfo {title} {Electromagnetic theory}}}\ (\bibinfo  {publisher} {Swedeborg Press},\ \bibinfo {year} {2010})\BibitemShut {NoStop}%
\end{thebibliography}%

\end{document}